\documentclass[pra, reprint, twocolumn, superscriptaddress, amsfonts, amssymb, amsmath]{revtex4-2}
\usepackage[english]{babel}

\usepackage{graphicx}

\graphicspath{{./Images/},{./ImagesAppendix/},
{./images/},{./imagesAppendix/}}

\usepackage{physics}
\usepackage{bm}
\usepackage{braket}
\usepackage{pgfplots}
\usepackage{array}
\usepackage{makecell}
\usepackage{amsmath, amssymb, amsthm, mathtools, xfrac}
\usepackage{tikz}
\usepackage{subfigure}
\usepackage{commath}
\usepackage{graphicx,bm}
\usepackage{verbatim}
\usepackage{flafter}
\usepackage{float}
\usepackage{varioref}
\usepackage{dsfont}

\usepackage{mathtools}

\usepackage{color}
\usepackage[colorlinks,citecolor=darkBlue,linkcolor=darkBlue,
urlcolor=blue,hyperindex]{hyperref}
\definecolor{darkBlue}{rgb}{0.08, 0.13, 0.4}

\definecolor{THc}{rgb}{0.9,0.3,0.2}

\newcommand{\canc}[1]{}

\begin{document}

\title{Local reminiscence in the PXP model}
\author{F. Perciavalle}
\affiliation{Dipartimento di Fisica, Universit\`a della Calabria, 87036 Arcavacata di Rende (CS), Italy}
\affiliation{INFN--Gruppo collegato di Cosenza}

\author{G. M. Rizzo}
\affiliation{Dipartimento di Fisica, Universit\`a della Calabria, 87036 Arcavacata di Rende (CS), Italy}

\author{F. Plastina}
\affiliation{Dipartimento di Fisica, Universit\`a della Calabria, 87036 Arcavacata di Rende (CS), Italy}
\affiliation{INFN--Gruppo collegato di Cosenza}

\author{N. Lo Gullo}
\affiliation{Dipartimento di Fisica, Universit\`a della Calabria, 87036 Arcavacata di Rende (CS), Italy}
\affiliation{INFN--Gruppo collegato di Cosenza}

\date{\today}

\begin{abstract}
We study the emergence of \textit{local reminiscence} in the PXP model, a constrained spin system realized in Rydberg atom arrays. The spectrum of this model is characterized by a majority of eigenstates that satisfy the eigenstate thermalization hypothesis, alongside a set of nonthermal eigenstates, known as quantum many-body scars, that violate it. We show that special configurations generate local reminiscent dynamics by preserving memory of the initial state. We explore local memory retention using local fidelity and the dynamics of local observables. We find that two specific states, $\theta$-symmetric and blockaded states, exhibit robust local reminiscence, with fidelities near unity and suppressed fluctuations as the system size increases. Our results show that non-ergodic regimes can sustain stable local memory while still allowing for complex global dynamics, providing new insights into quantum many-body scars and constrained dynamics.
\end{abstract}

\maketitle

\section{Introduction}
\begin{figure}[!t]
\centering
\includegraphics[width=0.78\linewidth]{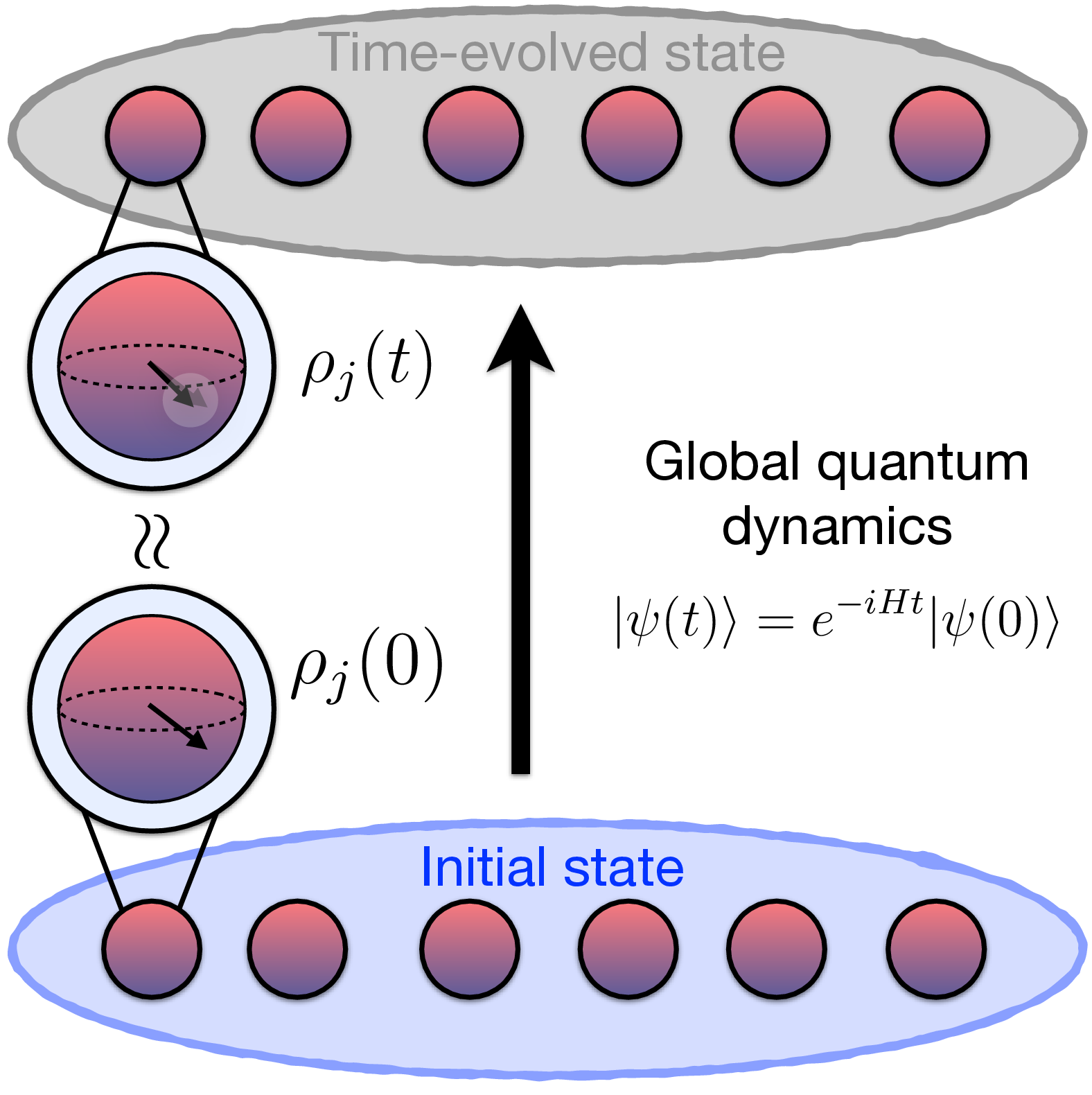}
\caption{Sketch of a local reminiscent dynamics: a system composed of $L$ qubits is initialized in the state $\ket{\psi(0)}$, whose $j$-site local density matrices 
$\rho_j(0)$ are obtained by tracing out the degrees of freedom of the rest of the chain from $\ket{\psi(0)}\bra{\psi(0)}$. The system undergoes a global time evolution governed by a Hamiltonian $H$, and this evolution is pictorially represented by a change in color of the global state. The system is locally reminiscent if local states, such as the time-evolved 
$j$-site local density matrices $\rho_j(t)$, are close to their counterparts at $t=0$.}
\label{fig:sketch_main}
\end{figure}
The study of thermalization in isolated many-body quantum systems has uncovered deep links between ergodicity, quantum chaos, and the dynamics of information. According to the eigenstate thermalization hypothesis (ETH)~\cite{rigol2008thermalization, srednicki1994chaos, deutsch1991quantum, deutsch2018eigenstate, alba2015eigenstate, mori2018thermalization}, non-integrable interacting quantum systems governed by the Schrödinger equation can undergo local thermalization by effectively erasing memory of their initial conditions. In such ergodic regimes, local observables relax to thermal equilibrium values, and the system's long-time evolution aligns with the predictions of statistical mechanics. Notably, entanglement entropy in these regimes exhibits extensive, volume-law scaling, reflecting the system's thermal nature~\cite{singh2016signatures, calabrese2005evolution, nandkishore2015many}.
 
Recent research, however, has uncovered surprising deviations from the established paradigm. Specifically, non-integrable quantum systems described by the PXP model~\cite{bernien2017probing,turner2018quantum, lesanovsky2012interacting, ho2019periodic, serbyn2021quantum, hudomal2022driving, liang2025observation, surace2021exact, park2025nonintegrability}, which can be realized on Rydberg atom platforms~\cite{browaeys2020many, wu2021concise, adams2019rydberg, saffman2010quantum, morgado2021quantum}, have displayed unexpected dynamical behavior when initialized in certain states. Despite their general non-integrable nature, these systems do not always thermalize as predicted by the ETH, thus challenging the conventional understanding of quantum ergodicity. A prominent example is the evolution of a Néel state in such systems, where, the system retains memory of its initial conditions. Rather than exhibiting thermalization, the dynamics of local observables are marked by regular oscillations. This phenomenon has been attributed to the presence of quantum many-body scars (QMBS)~\cite{serbyn2021quantum, chandran2023quantum, choi2019emergent, pappalardi2025theory, desaules2022extensive, kerschbaumer2025quantum, daniel2023bridging, pizzi2025genuine}, which are a small subset of special eigenstates that violate the ETH and prevent thermalization.

In this work, we investigate the dynamics of the system, with a particular focus on the preservation of local information. Specifically, we address the question: can a system retain memory at the local level, even when its global dynamics become complex or non-periodic? We refer to this phenomenon as \textit{local reminiscence}, that we identify as the stability of the reduced density matrices of small subsystems with respect to their initial configurations during the dynamics, the phenomenon is pictorially represented in Fig.~\ref{fig:sketch_main}. Detecting and characterizing this effect not only deepens our understanding of quantum many-body scarring but also sheds light on the mechanisms underlying local memory retention in constrained quantum systems. We mention that this phenomenon is potentially related to non-markovianity of the time-evolution of the local system, that prevents thermalization and has been recently studied in spin chains such as the quantum Ising model and the deformed PXP model~\cite{banerjee2025non, banerjee2025quantum}.

Motivated by recent theoretical and experimental advances in the control and preparation of quantum states on tunable platforms~\cite{balewski2024engineering,semeghini2021probing,bornet2024enhancing,haug2021machine,carrera2025preparing,briongosmerino2025dipolar,perciavalle2024quantum, lutz2025adiabatic}, in this paper, we systematically investigate local reminiscence in the PXP model, initialized in quantum states that deviate from the commonly studied configurations. The latter are high-energy states and show large overlaps with few QMBS at the edge of the spectrum. The investigation is conducted by analyzing the dynamics of local fidelities, which are the fidelities between reduced quantum states on small subsystems at time zero and after time evolution, as well as the dynamics of local observables. We first consider a probe state controlled by a parameter $\theta$, interpolating between the Néel state (non-ergodic dynamics) and the homogeneous state with all spins down (ergodic dynamics). The probe state itself does not exhibit local reminiscence, but its superposition with a spatially inverted symmetric state, the $\theta$-symmetric state, does. When initialized in this superposition, the system stably retains local memory of its initial condition, despite lacking global memory. Additionally, we introduce the blockaded state, a superposition of computational basis states with no neighboring excitations. Governed by the Fibonacci sequence, this state shows strong local reminiscence. Although their global dynamics differ, both the $\theta$-symmetric and blockaded states share common local reminiscent features.

The remainder of the paper is organized as follows: in Sec.~\ref{sec:modmet} we present the model Hamiltonian together with its main features of interest for this work. We also give a brief overview of scarred state in the context of ergodic dynamics and introduce the concept of local reminiscence. We introduce some quantities which will be used in the remainder of the paper. In Secs.~\ref{sec:theta},~\ref{sec:theta_sym},~\ref{sec:block} we introduce new many-body states which have peculiar dynamics; specifically we will show that although some of them behave similarly to the widely studied \(\ket{\mathbb{Z}_2}\) state when the global dynamics is considered, they have sharply different local features. In Sec.~\ref{sec:spectr} we discuss the spectral properties of the Hamiltonian and we show that the local reminiscence occurs for those initial states which have a larger overlap with the pure point spectrum of the system Hamiltonian. We also show that scarred many-body states are part of this spectrum. In Sec.~\ref{sec:concl} we draw the conclusions and give some perspectives on how our work can foster the study of states different from the \(\ket{\mathbb{Z}_2}\) one finding equally rich physics.
 
\section{Model and methods}
\label{sec:modmet}
We consider an array of $L$ qubits described by a PXP Hamiltonian with open boundary conditions (OBC)~\cite{turner2018quantum, lesanovsky2012interacting, ho2019periodic, serbyn2021quantum, hudomal2022driving}
\begin{equation}
    H_{\rm PXP} = \frac{\Omega}{2} \Bigl(X_1 P_2 + \sum_{j=2}^L P_{j-1}X_j P_{j+1} + P_{L-1}X_{L}\Bigr),
\end{equation}
where $X_{j}$ is the Pauli-$X$ operator acting on the $j$-th qubit, such that $X_{j}\ket{0}_j = \ket{1}_j$ and $X_{j}\ket{1}_j=\ket{0}_j$, while $P_j=\frac{1}{2}(1-Z_j)$ is the projector onto the $\ket{0}_j$ state, with $Z_j$ being the Pauli-$Z$ operator, satisfying $Z_j\ket{0}_j = -\ket{0}_j$ and $Z_j\ket{1}_j = \ket{1}_j$. The dynamics of a quantum system initialized in the state $\ket{\psi(0)}$ is described by the Schrödinger equation $\ket{\psi(t)}=\exp(-i H_{\rm PXP}t)\ket{\psi(0)}$, where we have fixed $\hbar=1$. We also fix $\frac{\Omega}{2}=1$, which means that the times are in units of $\bigl(\frac{\Omega}{2}\bigr)^{-1}$ and the energies are in units of $\frac{\Omega}{2}$. We consider a system composed of an even number of sites $L$. The PXP Hamiltonian describes a highly constrained system of Rydberg atoms where the interplay between coherent driving and interaction-induced constraints gives rise to rich and exotic dynamics~\cite{liang2025observation}. 
In particular, the PXP model is realized in the limit in which both the energy scales associated with Rabi frequency and detuning are much smaller than nearest-neighbor interaction~\cite{lesanovsky2011many, lesanovsky2012interacting, lesanovsky2012liquid, turner2018quantum, serbyn2021quantum}, more details are reported in appendix~\ref{app:details_derivation}. 

\subsubsection{Ergodic vs scarred dynamics}
The dynamics of the system is constrained to a subspace in which two or more nearest-neighbor excitations are prohibited, a condition referred to as a ``blockaded''. Quantitatively, the Hamiltonian commutes with the operator $\mathcal{P}=\prod_j(1 - n_j n_{j+1})$~\cite{omiya2023fractionalization}, which acts as the identity on states within the blockaded subspace (i.e., states without adjacent excitations) and annihilates any state with support outside this subspace. Therefore, the time evolution of an initial state belonging to the blockaded subspace remains entirely constrained within it. Thus, for this family of initial states, the effective dimension of the Hilbert space, which would normally consist of $2^L$ basis states, is reduced to a Fibonacci-scaling subspace of size $F_{L+2}$, where $F_n$ denotes the \( n \)-th Fibonacci number~\cite{bernien2017probing,turner2018quantum}.  

A large part of the spectrum is thermal, in the sense that it satisfies the ETH, and the dynamics that they generate is quantum chaotic (or ergodic), in the sense that local observables tend to relax at long times and the system quickly looses information on its initial configuration, i.e. the fidelity between the initial state and its time-evolved version $\mathcal{F}(t)=|\braket{\psi(0)|\psi(t)}|^2$ goes to zero. Another quantity whose dynamics signals the presence of ergodic behavior is the entanglement entropy 
\begin{align}
&S(t)=-\operatorname{Tr}_{\mathcal{S}}\bigl[\rho_{\mathcal{S}}(t)\ln \rho_{\mathcal{S}}(t)\bigr],
\label{eq:ententr}
\end{align}
where $\rho_{\mathcal{S}}(t)=\operatorname{Tr}_{\bar{\mathcal{S}}}\left[\ket{\psi(t)}\bra{\psi(t)} \right]$, $\mathcal{S}$ is a sub-part of the system and $\bar{\mathcal{S}}$ its complement. The entanglement entropy dynamics of an ergodic system is characterized by a linear growth at short times and then a saturation to a value that linearly depends on the size of the subsystem, reflecting the so-called volume-law of entanglement~\cite{dalessio2016from, calabrese2005evolution, amico2008entanglement, horodecki2009quantum}. In the case of the PXP model, an example of state whose dynamics is clearly chaotic is the homogeneous state $\ket{\boldsymbol{0}}=\ket{000000\ldots}$. This state has basically homogeneous overlap with the eigenstates of the Hamiltonian, thus it is uniformly distributed along the spectrum. 

However, when the system is initialized in the Néel state $\ket{\mathbb{Z}_2}=\ket{101010\ldots}$, or alternatively in $\ket{\mathbb{Z}_2^{'}}=\ket{010101\ldots}$, the dynamics exhibits strong departures from conventional thermalization behavior, showing coherent oscillations over long time scales in local observables, fidelity and entanglement entropy dynamics, instead of saturating. This type of behavior has been also experimentally observed in arrays composed of interacting Rydberg atoms~\cite{bernien2017probing, liang2025observation}. The phenomenon was attributed to the presence of \textit{quantum many-body scars} (QMBS), a small subset of anomalous eigenstates embedded within an otherwise thermal spectrum that exhibit unusually high overlap with specific initial configurations, such as the $\ket{\mathbb{Z}_2}$ state. It follows that initial configurations showing such an anomalous behavior have a large overlap with these scar states.

\subsubsection{Local reminiscence}
In this paper, we focus on exploring the local properties of the system's dynamics and how they depend on the choice of the initial state. To this end, we examine the relations between the reduced density matrices on specific subsystems $\mathcal{S}$ of the chain. In particular, we are interested in comparing the reduced density matrix of the initial state
\begin{equation}
    \rho_{\mathcal{S}}(0)=\operatorname{Tr}_{\bar{\mathcal{S}}}\bigl[\ket{\psi(0)}\bra{\psi(0)}\bigr]
\end{equation}
with that of the time-evolved state
\begin{align}
\rho_{\mathcal{S}}(t)=\operatorname{Tr}_{\bar{\mathcal{S}}}\bigl[e^{-iH_{\rm PXP} t}\ket{\psi(0)}\bra{\psi(0)}e^{i H_{\rm PXP} t}\bigr].
\end{align}
The main question that drives this analysis is whether there exist special initial conditions under which the system retains, in a stable manner, memory at the local level, that is, $\rho_{\mathcal{S}}(t) \approx \rho_{\mathcal{S}}(0)$, while still evolving non-trivially at the global scale. When referring to local scales, we consider cases where $\textrm{size}\,\mathcal{S} \ll L$, meaning reduced density matrices on single sites or few-site subsystems within a finite system of size $L$. We refer to this type of dynamics as \textit{local reminiscent} and we pictorially represent it in Fig.~\ref{fig:sketch_main}.  A paradigmatic example of states that satisfy this type of properties is given by the entangled Greenberger–Horne–Zeilinger (GHZ)~\cite{greenberger1989going} states $\ket{\mathrm{GHZ}_+} = \frac{1}{\sqrt{2}}(\ket{\boldsymbol{0}} + \ket{\boldsymbol{1}})$ and $\ket{\mathrm{GHZ}_-} = \frac{1}{\sqrt{2}}(\ket{\boldsymbol{0}} - \ket{\boldsymbol{1}})$, where $\ket{\boldsymbol{1}} = \ket{111\ldots}$.
They are globally orthogonal but locally indistinguishable. In fact, all of their single-site reduced density matrices are identical, given by $\frac{1}{2} \mathds{1}$, making them locally indistinguishable.

We study local reminiscence by using as quantifier the Uhlmann fidelity between the reduced initial and time evolved states~\cite{nielsen2012quantum,uhlmann1976transition}
\begin{equation}
    \mathcal{F}_\mathcal{S}(t) = \mathcal{F}(\rho_\mathcal{S}(0), \rho_\mathcal{S}(t)) \equiv \Tr \left[ \sqrt{\sqrt{\rho_\mathcal{S}(0)}\rho_\mathcal{S}(t)\sqrt{\rho_\mathcal{S}(0)}} \right]^2;
    \label{eq:F_s}
\end{equation}
In particular, in this paper we focus our analysis on single-site reduced density matrices $\rho_{j}(t)$, obtained by tracing out all sites of the chain except site $j$ from the full state $\ket{\psi(t)}\bra{\psi(t)}$. Thus, the associated $j$-site local fidelity is defined as $\mathcal{F}_j(t)$. The local fidelity is directly related to the trace distance between the two reduced density matrices $\mathcal{D}_j(t)=\frac{1}{2}\|\rho_j(t) - \rho_j(0) \|_1$, through the following bounds: $1-\sqrt{\mathcal{F}_j(t)}\leq \mathcal{D}_j(t)\leq \sqrt{1 - \mathcal{F}_j(t)}$~\cite{nielsen2012quantum}. These inequalities establish a lower and upper bound on the trace distance in terms of the fidelity, for this reason, we introduce $\mathcal{D}_{j}^{\rm max}(t)\coloneqq \sqrt{1 - \mathcal{F}_j(t)}$, which represents the maximum possible distance between the evolved $j$-site reduced density matrix and its initial counterpart at $t=0$.

For exploring local reminiscence, we monitor the dynamics of the following quantities:
\begin{itemize}
\item The product of all the single-site local fidelities $\prod_{j=1}^L\mathcal{F}_j(t)$, that corresponds to the global fidelity in the case in which the the time-evolved state is factorized, $\rho(t)=\prod_{j=1}^{L}\rho_j(t)$. This object is of interest for two main reasons: first, it allows quantifying in one single shot the effect of all the single-site local fidelities, it is sufficient to have a single small local fidelity to have a small overall contribution. Secondly, this expression captures how local fidelities $\mathcal{F}_j(t)$ would combine in the absence of correlations, and thus serves as a meaningful reference to be compared with the actual global fidelity $\mathcal{F}(t)$. Indeed, the inequality $\prod_{j=1}^{L}\mathcal{F}_j(t)>\mathcal{F}(t)$ signals that the collective effect of the local fidelities exceeds the actual global fidelity, indicating that while local memory is preserved, correlations between sites hinder the recovery of the global state.
\item The spatial average of the local fidelity 
    \begin{equation}
    \mathcal{F}_{1\rm-site}(t)=\frac{1}{L}\sum_{j=1}^L \mathcal{F}_j(t).
     \end{equation}
      that quantifies the average behavior on the whole chain of the local single-site fidelities. In this case, a single small local fidelity does not necessarily lead to a small overall contribution of $\mathcal{F}_{1\rm-site}(t)$; thus, it remains basically large if the majority of sites have large local fidelities.
\item The time-averaged local fidelity (Césaro average)
\begin{equation}
    \bigl(\overline{\mathcal{F}}_{\rm 1-site} \bigr)_t = \frac{1}{t} \int_0^t  \mathcal{F}_{\rm 1-site}(t') \, dt',
    \label{eq:time-avg}
\end{equation}
and its associated standard deviation
\begin{equation}
    \bigl( \sigma_{\mathcal{F}_{1-\textrm{site}}} \bigr)_t = \sqrt{\frac{1}{t} \int_{0}^t \Bigl[\mathcal{F}_{\rm 1-site}(t') - \bigl(\overline{\mathcal{F}}_{\rm 1-site}\bigr)_{t'} \Bigr]^2  dt' }.
    \label{eq:stddev}
\end{equation}
They quantify magnitude and fluctuations of the signal represented by the local fidelity, and thus the stability of the local reminiscence. In few words, having $    \bigl(\overline{\mathcal{F}}_{\rm 1-site} \bigr)_t\rightarrow 1$ and $\bigl( \sigma_{\mathcal{F}_{1-\textrm{site}}} \bigr)_t \rightarrow 0$ in the presence of global evolution indicates strong and stable local reminiscence.
\end{itemize}
In the following, we analyze these dynamical features for different initial states, showing that there are particular configurations for which a local reminiscent dynamics clearly emerges.

\section{The $\ket{\Theta_+}$ state}
\label{sec:theta}
We consider probe states that interpolate between homogeneous and Néel states, allowing us to explore both global and local dynamics across different regimes. Specifically, we introduce the $\ket{\Theta_+}$-family of states, which is defined by the states
\begin{align}
    &\ket{\Theta_+}=\ket{\theta_+}\otimes\ket{0}\otimes\ket{\theta_+}\otimes\ket{0}\otimes\ldots,
\\ & \ket{\Theta_+^{'}}=\ket{0}\otimes\ket{\theta_+}\otimes\ket{0}\otimes\ket{\theta_+}\otimes\ldots,
\end{align}
where $\ket{\theta_+}\coloneqq \cos(\theta)\ket{0} +\sin(\theta)\ket{1}$. These two states interpolate the homogeneous state $\ket{\boldsymbol{0}}$, obtained for $\theta=0$, and the Néel states $\ket{\mathbb{Z}_2}$ and $\ket{\mathbb{Z}_2^{'}}$, obtained for $\theta=\pi/2$; thus, the mixing angle $\theta$ is a tunable parameter that allows us to probe the crossover between ergodic and scarred regimes, traversing an intermediate region.
\begin{figure}[!t]
\centering
\includegraphics[width=\linewidth]{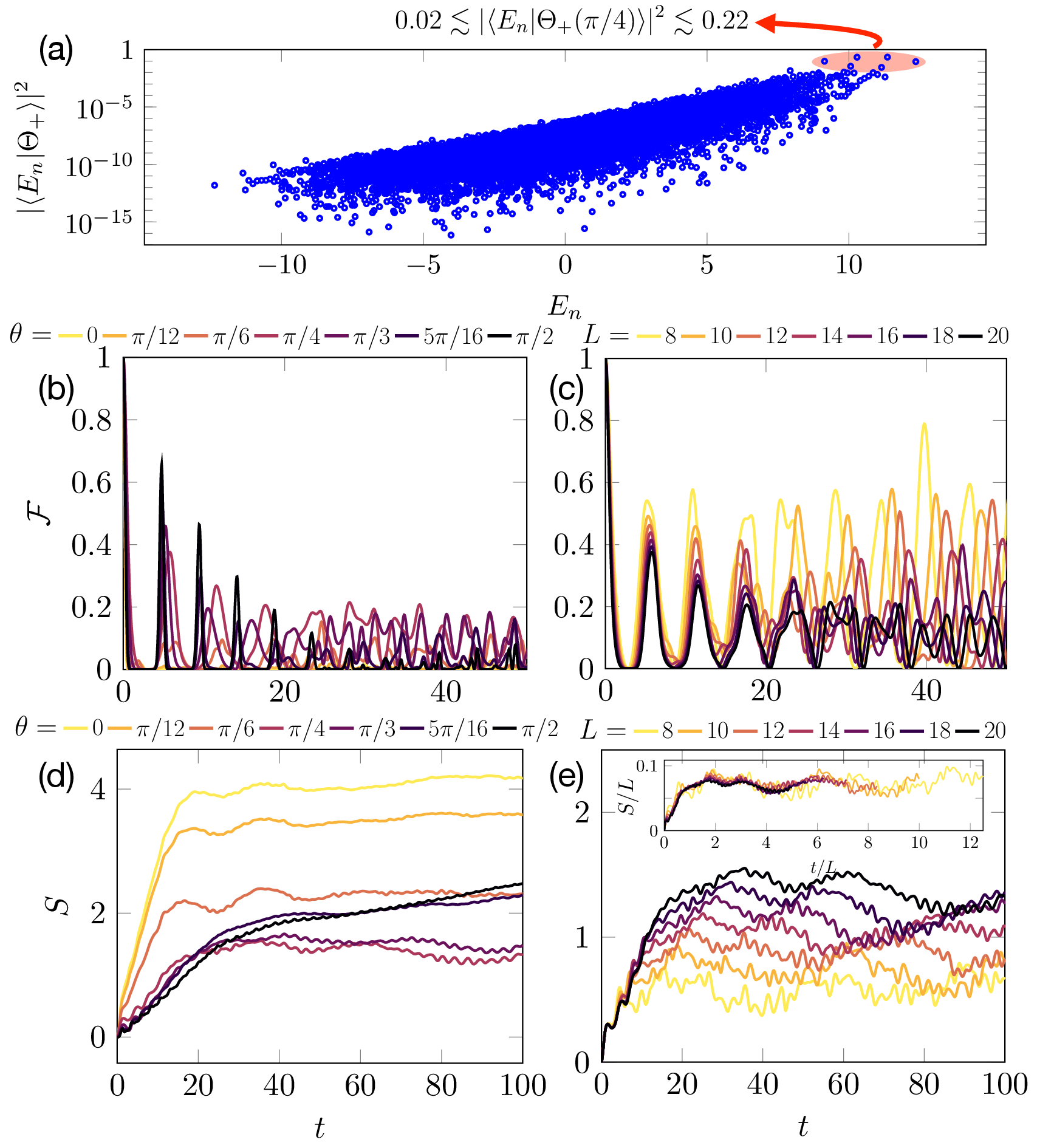}
\caption{Global properties of the dynamics of the $\ket{\Theta_+}$ state. Panel (a): overlap of the state with the eigenstates of the Hamiltonian for $\theta=\pi/4$ and $L=20$. The states with larger overlap are highlighted in red. Panels (b),(c): dynamics of the global fidelity for $L=20$ and different values of $\theta$ (b) and for $\theta=\pi/4$ and different values of $L$. Panels (d),(e): dynamics of entanglement entropy for $L=20$ and different values of $\theta$ (d) and for $\theta=\pi/4$ and different values of $L$ (e), the inset reports the latter with entanglement entropy and times rescaled by $L$.}
\label{fig:sketch_theta+}
\end{figure}

We first analyze the overlap of the $\ket{\Theta_+}$ state with the eigenstates of the Hamiltonian. For $\theta=0$, the overlap is uniformly distributed across eigenstates, while, for $\theta=\pi/2$, it is significantly enhanced in correspondence of the QMBS states. In Fig.~\ref{fig:sketch_theta+}(a), we report the overlap for the intermediate case $\theta=\pi/4$ and $L=20$. The behavior of the overlap in the spectrum in semi-log axis is linear, indicating a fast exponential growth of the latter with the energy of the system. Thus, differently from both $\theta=0$ (homogeneous state) and $\theta=\pi/2$ (Néel state), the state is highly influenced from a small group of high-energy states. In the figure we highlight the states with the larger overlap, that approximately ranges from $0.02$ to $0.22$.

Secondly, we analyze the global fidelity between the initial state $\ket{\psi(0)}=\ket{\Theta_+}$ and the time-evolved state $\ket{\psi(t)}$, $\mathcal{F}(t)$, which quantifies the system’s ability to retain or lose memory of its initial conditions at the global level. In Fig.~\ref{fig:sketch_theta+}(b), we report the dynamics of the aforementioned quantity for fixed $L=20$ and different values of the mixing angle $\theta$, in order to explore its behavior in the ergodic-scarred crossover. As the parameter $\theta$ is varied from $\theta = \frac{\pi}{2}$ towards $\theta = 0$, the characteristic revivals observed in the dynamics gradually disappear. In the intermediate regime, particularly around $\theta \approx \frac{\pi}{4}$, the system exhibits a noticeable loss of regularity in its temporal evolution, signaling a transition away from the regular revival pattern. 
In the latter regime, the state has large overlap with the high-energy eigenstates of the system, see Fig.~\ref{fig:sketch_theta+}(a). For this reason, in Fig.~\ref{fig:sketch_theta+}(b), we observe that, even at large times, the fidelity remains consistently greater than zero. 
In Fig.~\ref{fig:sketch_theta+}(c) we fix the mixing angle to the intermediate value $\theta=\pi/4$, and analyze the behavior of the fidelity for different values of the size $L$. The resulting behavior deviates from the traditional quantum many-body scarring observed in the dynamics of the $\ket{\mathbb{Z}_2}$ state. At short times, the system exhibits fidelity revivals characterized by distinct peaks, whose magnitude decreases with increasing system size $L$. However, in contrast to the dynamics initiated from $\ket{\mathbb{Z}_2}$, the long-time behavior becomes highly irregular and shows a pronounced dependence on $L$. 

Subsequently, we analyze the entanglement entropy $S(t)$, defined in Eq.~\eqref{eq:ententr}, that we compute by considering the subsystem $\mathcal{S}$ to be left half of the chain, comprising sites $[1, L/2]$ (with $L$ even), and $\bar{\mathcal{S}}$ as the right half, comprising sites $[L/2 + 1, L]$. In Fig.~\ref{fig:sketch_theta+}(e), we report the dynamics for fixed $L = 20$ and different values of $\theta$. For $\theta = 0$, the dynamics is ergodic, as evidenced by a linear growth of the entanglement entropy followed by saturation to a volume-law plateau~\cite{papic2022weak}. As $\theta$ increases towards $\frac{\pi}{2}$, this behavior gradually changes: the entropy growth becomes associated with scarred dynamics, which is non-ergodic and does not exhibit saturation to a plateau. In the intermediate regime, the entanglement entropy reaches values that are comparable in magnitude to those observed in the non-ergodic, scarred regime. However, its temporal behavior more closely resembles that of the ergodic regime, displaying a feature akin to a plateau. Unlike the ergodic case, though, this plateau is not stationary but exhibits oscillations over time. Its scaling with the size can be appreciated in Fig.~\ref{fig:sketch_theta+}(f), in which the entanglement dynamics is reported for fixed $\theta=\pi/4$ and varying system sizes $L$. As the system size increases, the entanglement entropy also increases. As shown in the inset, rescaling both time and entanglement entropy by the system size $L$ leads to a data collapse, indicating a scaling behavior of the form $S(t, L) \sim L \, f\left(\frac{t}{L}\right)$. This suggests a volume-law growth of entanglement. However, unlike in ergodic quantum systems, where the entropy saturates to a thermal plateau at long times, here the entropy continues to exhibit oscillations. This suggests that, although the dynamics exhibits certain characteristics typically associated with ergodic behavior, it cannot yet be classified as ergodic. At the system sizes accessible in this study, the evidence remains insufficient to unambiguously establish ergodicity.
\subsubsection{Local features of the dynamics}
\begin{figure}[!t]
\centering
\includegraphics[width=\linewidth]{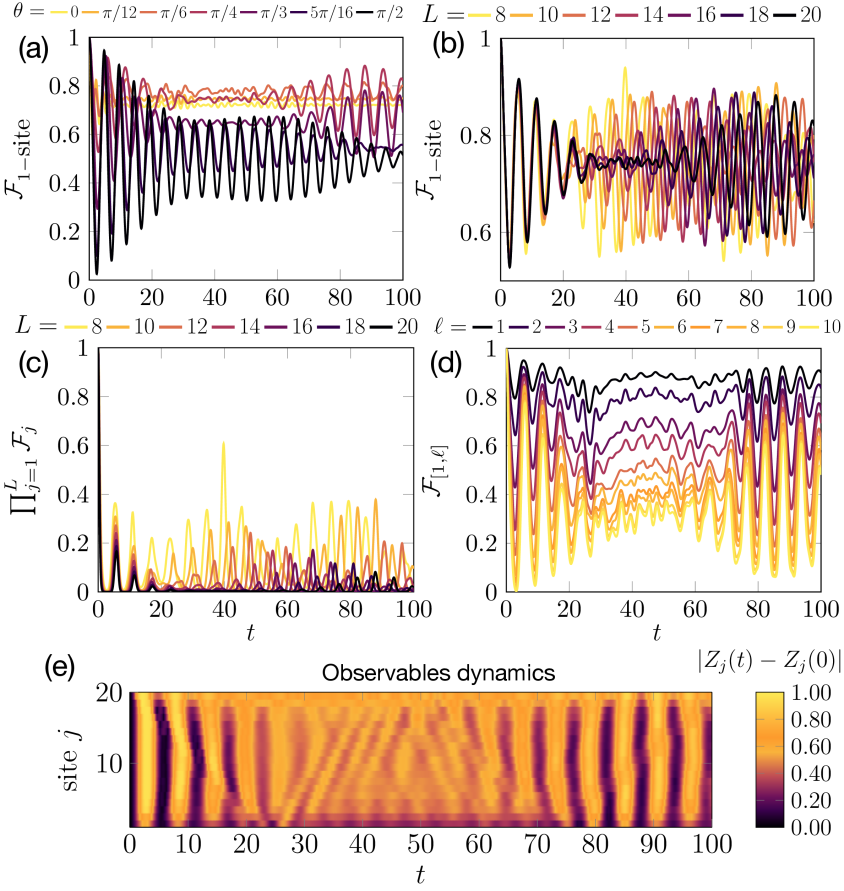}
\caption{Local properties of the dynamics of the $\ket{\Theta_+}$ state. Panel (a) and (b): dynamics of the single-site local fidelity averaged over the whole chain for $L=20$ and different $\theta$ (a), $\theta=\pi/4$ and different $L$(b). Panel (c): product of the all the single-site local fidelities for different values of $L$ and $\theta=\pi/4$. Panel (d): local fidelity evaluated on blocks $\mathcal{S}=[1,\ell]$ of different size with fixed $\theta=\pi/4$ and $L=20$. Panel (e): dynamics fluctuation of the $Z$-magnetization with respect to its initial value in each site for $L=20$ and $\theta=\pi/4$.}
\label{fig:local_theta+}
\end{figure}
We are now interested in understanding the behavior of this state at the local level by using the local fidelity $\mathcal{F}_\mathcal{S}(t)$ between the initial and the time-evolved reduced density matrices, searching for \textit{local reminiscence}. We investigate local memory behavior in the crossover from ergodic ($\theta = 0$) to scarred ($\theta = \pi/2$) dynamics. As subsystem $\mathcal{S}$, we can consider either a single site $j$ or a finite block of sites $[1, \ell]$. We refer to the fidelity in Eq.~\eqref{eq:F_s} as local when $\ell \ll L$. 

In Figs.~\ref{fig:local_theta+}(a) and~\ref{fig:local_theta+}(b) we examine the dynamical behavior of $\mathcal{F}_{1-\rm site}(t)$. Fig.~\ref{fig:local_theta+}(a) reports the analysis of the dynamics for fixed $L=20$ and varying $\theta$. The two asymptotic values of $\theta$, $\theta=0$ and $\theta=\pi/2$, exhibit distinctly different behaviors. For $\theta=0$, the system approximately relaxes to a value significantly lower than one, with small oscillations around this value; while, for $\theta=\pi/2$, as in the case of the global dynamics, it undergoes strong oscillations that, in magnitude, are much smaller than those observed for $\theta=0$. Thus, while the system initialized in $\ket{\boldsymbol{0}}$ rapidly loses its global memory and thermalizes, its local behavior is markedly different: the local equilibrium state remains unexpectedly close to the initial one, even more so than in the Nèel case. In an idealized scenario where the Nèel state coherently oscillates between $\ket{\mathbb{Z}_2}$ and $\ket{\mathbb{Z}_2'}$, the local state would alternate between $\ket{0}_j$ and $\ket{1}_j$, causing the local fidelity to swing between $1$ and $0$; bulk effects and unequal level spacings reduce this to oscillations around $\sim0.5$. By contrast, this result highlights that the homogeneous state explores a much smaller portion of the local Hilbert space, which keeps its local fidelity significantly higher. As $\theta$ varies, the dynamics undergo a crossover between these two regimes, which is more clearly captured by the local fidelity than by its global counterpart. In fact, as $\theta$ increases, we observe how the stable behavior at $\theta=0$ gradually transitions into oscillatory dynamics, with $\theta=\pi/4$ representing the point where the global fidelity dynamics is a mixture of both asymptotic regimes. We define $(\overline{\mathcal{D}_{1\textrm{-site}}^{\rm max}})_t = \frac{1}{t} \int_0^t \left( \frac{1}{L} \sum_{j=1}^L \mathcal{D}_j^{\rm max}(t') \right) dt'$, the time average of the maximum trace distance between the initial and evolved $j$-site reduced states, averaged over the chain. At long times ($t = 100$), we obtain $(\overline{\mathcal{D}_{1\text{-site}}^{\rm max}})_{t=100} \approx 0.525,\, 0.496,\, 0.681$ for the initial states $\ket{\boldsymbol{0}},\, \ket{\Theta_+(\pi/4)},\, \ket{\mathbb{Z}_2}$, respectively. For comparison, we note that the trace distance between two states far in the Bloch sphere such as $\ket{0}$ and $\frac{1}{\sqrt{2}}(\ket{0} + \ket{1})$ is $0.5$; thus, during this dynamics, the local state remains substantially different from the initial one even at late times. In Fig.~\ref{fig:local_theta+}(b), we study the dynamics of $\mathcal{F}_{1-\rm site}$ for different values of $L$, observing that it remains strongly oscillatory and erratic for all the values of $L$. The difference between system sizes is characterized by an apparent stabilization of the oscillations at intermediate time scales, with this stabilization lasting longer for larger $L$. The latter, common to all system sizes, reflects the behavior in the thermodynamic limit. Oscillations restart when finite-size effects emerge, and so depending on $L$.

Secondly, in Fig.~\ref{fig:local_theta+}(c), we analyze the product of all the single-site local fidelities $\prod_{j=1}^L\mathcal{F}_j(t)$. 
We explore its dynamics for $\theta=\pi/4$ and varying the size of the system. For increasing $L$, we observe that the product of the local fidelities becomes increasingly suppressed, signifying that the influence of local memory diminishes at larger scales. The decay of this quantity is exponential in system size (see appendix~\ref{app:scaling_prod}), similarly to what one would expect for a factorized system composed of independent local states $\rho_j(t)$. In that case, the global fidelity tends to zero as $L$ increases.

We are also interested in understanding how increasing the size of the blocks on which the reduced state is computed affects the fidelity $\mathcal{F}_{[1,\ell]}(t)$. Thus, we explore its dynamics ranging from the local regime ($\ell \ll L$) to the global regime, where $\ell$ becomes comparable to $L$; the analysis is performed for $L=20$ and $\theta=\pi/4$, see Fig.~\ref{fig:local_theta+}(d). Increasing the size of the block corresponds to enlarging the Hilbert space of the reduced state, which results in a lower overall fidelity magnitude. The main difference between the cases is that the apparent stabilization of the dynamics lasts for a shorter duration as $\ell$ increases. Larger $\ell$ values correspond to larger oscillations of the fidelity, with smaller mean values.

Finally, we look at how these features of the dynamics affect local observables, focusing in particular on the local magnetization. To do this, we study the time evolution of local magnetization relative to its initial value. This is measured by the quantity $|Z_j(t) - Z_j(0)|$, with $Z_j(t)=\braket{\psi(t)|Z_j|\psi(t)}$, which is shown in Fig.~\ref{fig:local_theta+}(e) for system size $L = 20$ and angle $\theta = \pi/4$. We find that this quantity shows a pattern of regular  oscillations, which are occasionally interrupted by periods where the magnetization seems to stabilize for some time. This behavior is similar to what we observed in the local fidelity. The magnitude of the fluctuations is quite large, which is what we expect in a system where the dynamics is not local reminiscent.
\section{The symmetric $\ket{\Theta_+^{\rm symm}}$ state}
\label{sec:theta_sym}
\begin{figure}[!t]
\centering
\includegraphics[width=\linewidth]{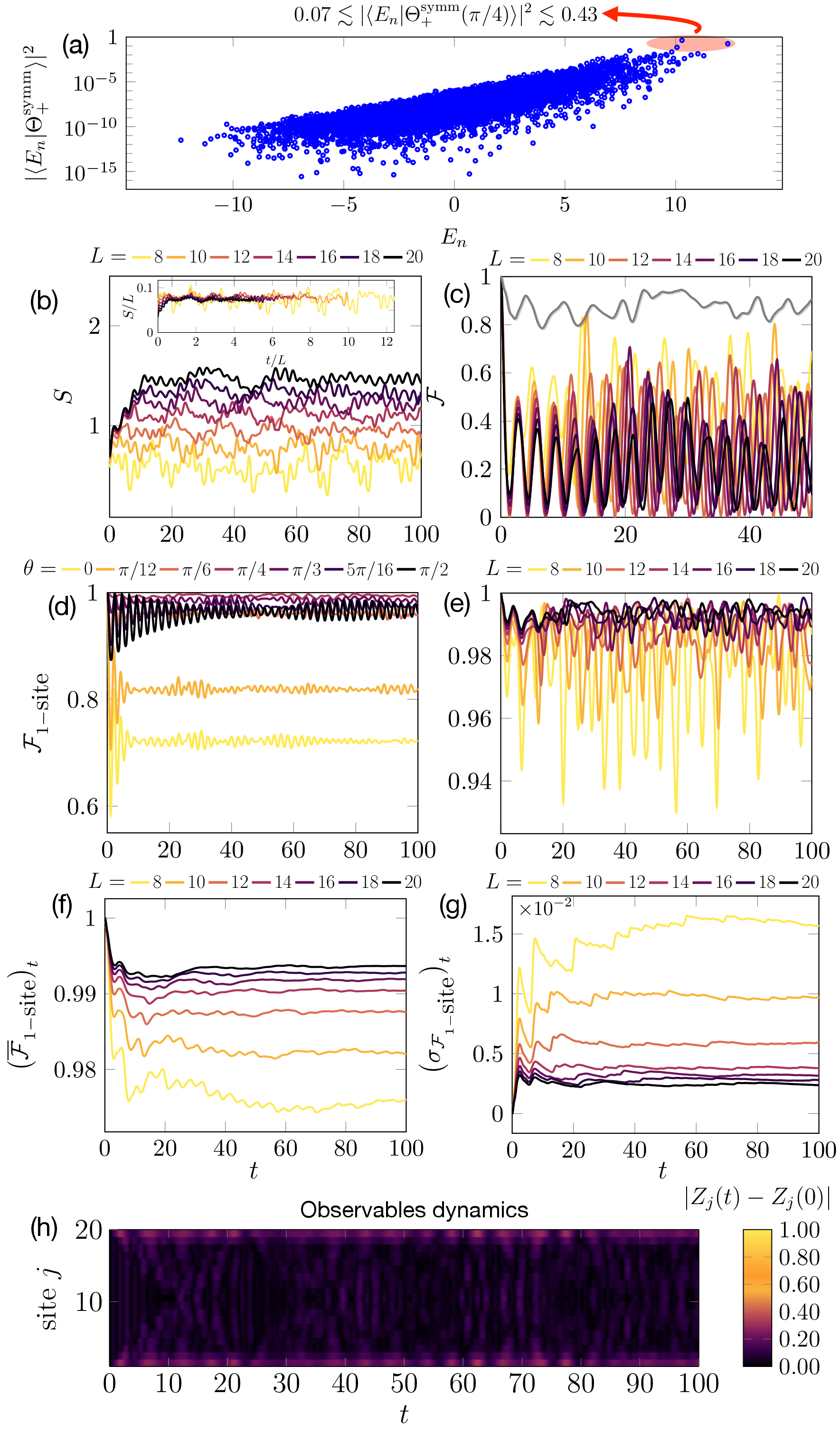}
\caption{Characterization of the $\theta$-symmetric state $\ket{\Theta_+^{\rm symm}}$ through its global and local features. Panels (a), (b) and (c) show the global features of the state, including the overlap with the eigenstates of the Hamiltonian for $L=20$ and $\theta=\pi/4$ (the states with larger overlap are highlighted in red) (a), the entanglement entropy and the global fidelity dynamics, respectively in panels (b) and (c), for $\theta=\pi/4$ and different $L$. In panel (b), the inset reports the dynamics of $S(t)$ for $\theta=\pi/4$ with both entanglement entropy and times rescaled by $L$. In panel (c), we compare the global fidelity with $\prod_{j=1}^L\mathcal{F}_j(t)$ for $L=20$ and $\theta=\pi/4$, which is reported in gray. Panels (d) and (e) show the dynamics of the local fidelity averaged over the whole chain, $\mathcal{F}_{1-\rm site}$. The dynamics is reported for $L=20$ and different values of $\theta$ (d), and for $\theta=\pi/4$ and different $L$ (e). Panels (f) and (g) show the time-average of the local fidelity $\mathcal{F}_{1-\rm site}$ and its associated standard deviation for $\theta=\pi/4$ and different $L$.  Panel (h) shows dynamics fluctuation of the $Z$-magnetization with respect to its initial value in each site for $\theta=\pi/4$ and $L=20$.}
\label{fig:sketch_ent}
\end{figure}
In this section, we consider an initial state built as the superposition of $\ket{\Theta_+}$ and $\ket{\Theta_+^{'}}$ states
\begin{equation}
    \ket{\Theta_+^{\rm symm}}=\frac{\ket{\Theta_+}+\ket{\Theta_+^{'}}}{\sqrt{2(1+\cos^{L}(\theta))}},
\end{equation}
where the factor $\sqrt{2(1+\cos^{L}(\theta))}$, with $\cos^{L}(\theta)=\braket{0|\theta_+}^L=\braket{\Theta_+|\Theta_+^{'}}$, ensures proper normalization of the state for any $\theta$. For $\theta \rightarrow 0$, we have $\ket{\Theta_{+}^{\rm symm}}\rightarrow\ket{\boldsymbol{0}}$, while for $\theta \rightarrow \pi/2$, we have $\ket{\Theta_{+}^{\rm symm}}\rightarrow\frac{\ket{\mathbb{Z}_2}+\ket{\mathbb{Z}_2^{'}}}{\sqrt{2}}$. We name this state as $\theta$-symmetric, because it is invariant under the spatial inversion symmetry $j \rightarrow L+1 -j$. This contrasts with $\ket{\Theta_+}$ and $\ket{\Theta_+^{'}}$, which do not exhibit this symmetry when $L$ is even. In fact, for even $L$, $\ket{\Theta_+}$ maps to $\ket{\Theta_+^{'}}$, and vice versa. Moreover, differently from the single $\ket{\Theta_+}$ and $\ket{\Theta_+^{'}}$ states, this state contains a greater variety of blockaded configurations. For instance, for $\theta=\pi/4$, the state becomes a superposition of all the computational basis states that do not contain two or more neighbor excitations, with the additional constraint that, in each configuration, excitations are restricted to sites of the same parity. Thus, for example, for $L=4$, configurations of the type $\ket{0101}$ are allowed, whereas configurations of the type $\ket{1001}$ are not. As we will see in the following, the dynamics of the symmetric state contrast sharply with those of the factorized $\ket{\Theta_+}$, with local reminiscence emerging.

In Fig.~\ref{fig:sketch_ent}(a) we report the overlap of the entangled state with the spectrum of the system, for fixed $L=20$ and $\theta=\pi/4$ and compare it with the product of local fidelities. As in the case of $\ket{\Theta_+}$, the overlap of $\ket{\Theta_+^{\rm symm}}$ with the eigenstates grows exponentially with the energy, and only few states at high energy dominate. Differently from the previous case, the maximum overlap is $|\braket{E_{\bar{n}}|\Theta_+^{\rm symm}(\pi/4)}|^2 \approx 0.43$; in general, the number of eigenstates contributing significantly to the initial state is reduced. This slightly different structure leads to strikingly different effects on the local dynamics of the system, with the clear emergence of local reminiscence.

Before analyzing the local features of the dynamics, we first characterize the global behavior of the state by examining the evolution of the entanglement entropy and global fidelity at the intermediate value of the mixing angle $\theta=\pi/4$, respectively reported in Fig.~\ref{fig:sketch_ent}(b) and Fig.~\ref{fig:sketch_ent}(c). Regarding the entanglement entropy, we observe an initial imperfect linear growth, as in the case of $\ket{\Theta_+(\pi/4)}$, followed by a pseudo-plateau, where the entropy ceases to grow on average but exhibits large fluctuations. The inset shows the entanglement dynamics with both entropy and time rescaled by $L$. At short times, during the initial growth, the data do not collapse, indicating that the behavior is not volume-law in this regime. At later times, around the pseudo-plateau, the data collapse well, and the volume-law scaling $S(t,L) \sim L \, f\left( \frac{t}{L} \right)$ accurately describes the dynamics. In terms of magnitude, $S(t)$ is small. The global fidelity exhibits an erratic behavior without a clear dependence on $L$. This quantity is clearly non-ergodic, as it displays large revivals and is not significantly suppressed over time. We also report in gray the product $\prod_{j=1}^{L} \mathcal{F}_j(t)$ for $L=20$ and $\theta=\pi/4$, which dominates over the global fidelity. Intriguingly, examining the scaling of this quantity (see appendix~\ref{app:scaling_prod}), we observe that the product fidelity increases with $L$ and tends to saturate, indicating a pronounced robustness of the latter quantity. This indicates that memory is much better retained at the local level, suggesting that this state exhibits a locally reminiscent dynamics. 

The presence of local reminiscence can be better explored by analyzing the dynamics of the average on the whole chain of the local fidelity, $\mathcal{F}_{1-\rm site}(t)$. In Figs.~\ref{fig:sketch_ent}(d),(e) we report the latter for fixed $L=20$ and varying $\theta$ (d), and for fixed $\theta=\pi/4$ and varying $L$ (e). While for small values of $\theta$ the local fidelity exhibits the expected behavior, characterized by short-time oscillations and long-time saturation to a value significantly below one, increasing $\theta$ leads to a thickening of the local fidelity around values close to one, with the dynamics that becomes locally reminiscent. In particular, the intermediate value of the mixing angle, $\theta = \pi/4$, appears to be optimal in this respect. As done for $\ket{\Theta_+}$, we compute the average value of the upper bound on the trace distance for $\theta = \pi/4$, which at long times yields $(\overline{\mathcal{D}_{1\text{-site}}^{\rm max}})_{t=100} \approx 0.065$, indicating that the time-evolved state remains locally close to its initial configuration. We also note that for $\theta = \pi/2$, i.e., for the initial state $\frac{\ket{\mathbb{Z}_2} + \ket{\mathbb{Z}_2'}}{\sqrt{2}}$, the dynamics exhibits oscillations characteristic of the scarred dynamics of $\ket{\mathbb{Z}_2}$ (and $\ket{\mathbb{Z}_2'}$). However, compared to Fig.~\ref{fig:local_theta+}(a), these oscillations have a higher frequency and are more tightly concentrated around one, indicating improved preservation of local memory. These feature arises from symmetry. The scar states alternate in inversion quantum number, so an initial state with even inversion symmetry such as $\frac{\ket{\mathbb{Z}_2} + \ket{\mathbb{Z}_2'}}{\sqrt{2}}$ only overlaps with every second scar. As a result, the energy spacing is doubled, leading to oscillations at twice the frequency. In panel (e), we focus on the effect of system size for the optimal case $\theta = \pi/4$, showing that local reminiscence is enhanced at larger sizes, as its magnitude increases and its fluctuations decrease with $L$.

The scaling in size of magnitude and fluctuations of the local fidelity are well captured by the time-averaged local fidelity $\bigl(\overline{\mathcal{F}}_{\rm 1-site} \bigr)_t $ and its associated standard deviation $\bigl( \sigma_{\mathcal{F}_{1-\textrm{site}}} \bigr)_t$, respectively defined in Eq.~\eqref{eq:time-avg} and Eq.~\eqref{eq:stddev}. As mentioned above, we use them to quantify how strong and stable the local reminiscence of this dynamics is. In Fig.~\ref{fig:sketch_ent}(f), we report the time average of the local fidelity for fixed $\theta = \pi/4$ and varying $L$. As expected, increasing the system size leads to an increase in $\bigl(\overline{\mathcal{F}}_{\rm 1\text{-}site} \bigr)_t$, which tends to collapse to a value above $0.99$, sufficiently close to one. At the same time, its standard deviation, shown in Fig.~\ref{fig:sketch_ent}(g) for $\theta = \pi/4$ and varying $L$, displays the opposite behavior, collapsing to a value well below $0.05$, sufficiently close to zero. Both behaviors are indicators of strong and stable local reminiscence.

We also mention that under periodic boundary conditions, the one-site fidelity remains significant throughout the evolution and grows with system size, indicating a robust and increasingly strong form of local reminiscence (see appendix~\ref{app:pbc}). Its magnitude is consistently larger than in the OBC case considered here, showing that closing the boundaries enhances the preservation of local information by eliminating edge effects. Overall, PBC not only maintain but clearly reinforce the local reminiscent behavior of the state $\ket{\Theta^{\rm symm}_+(\pi/4)}$.

Finally, in Fig.~\ref{fig:sketch_ent}(h), we show the time evolution of the deviation of the local magnetizations relative to their initial values, quantified by $|Z_j(t) - Z_j(0)|$, for a system of size $L = 20$ and fixed angle $\theta = \pi/4$. Unlike the case of the state $\ket{\Theta_+(\pi/4)}$, the fluctuations are now significantly reduced. While some dynamics are still visible near the edges of the chain, in the central region we find that $Z_j(t) \approx Z_j(0)$ at all times. This indicates that, at the level of local observables, the system retains a strong memory of its initial state, a clear signature of local reminiscence.
\section{The blockaded state}
\label{sec:block}
In this section, we introduce and study the dynamics of the blockaded state, i.e. the equal-weight superposition of all computational basis states allowed by the Rydberg blockade constraint. For an $L$ sites 1D-chain, $\ket{\varphi_L}$ is defined as follows:
\begin{equation}
\ket{\varphi_L} = \frac{1}{\sqrt{F_{L+2}}} 
\sum_{\ket{\boldsymbol{s}} \in \mathcal{B}_L} \ket{\boldsymbol{s}},
\end{equation}
where $F_{L+2}=\dim(\mathcal{B}_L)$ is the dimension of the constrained Hilbert space $\mathcal{B}_L$. We remark that $\ket{\boldsymbol{s}}=\ket{s_1,s_2,\ldots,s_L}$ are states belonging to the computational basis $\mathcal{B}_L$ that do not contain two or more neighbor excitations. Unlike the $\theta$-symmetric state at $\theta = \pi/4$ discussed in the previous section, this state includes all computational basis states that do not contain two or more adjacent excitations, including those with excitations on sites of different parity. 
We also note that this state is maximally delocalized in the computational basis.

\begin{figure}[!t]
\centering
\includegraphics[width=\linewidth]{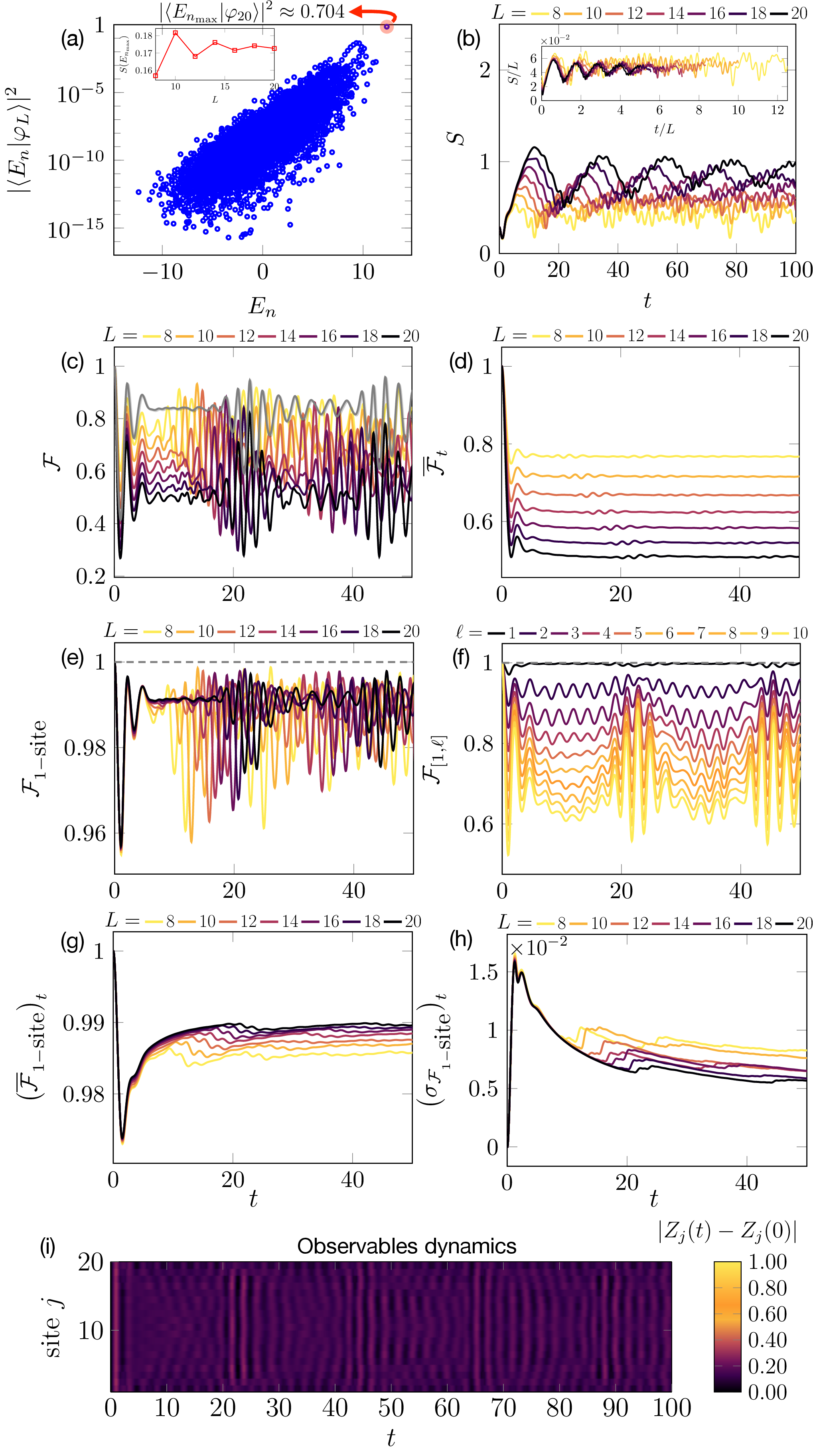}
\caption{Characterization of the blockaded state $\ket{\varphi_L}$. Panels (a),(b),(c) and (d) report the global features of the states: overlap with the eigenstates with fixed $L=20$ (a), the state with larger overlap is the highest excited state $\ket{E_{n_{\rm max}}}$ and it is highlighted in red; in the inset we report the scaling of its entanglement entropy with the size $L$. Panel (b) reports the dynamics of the entanglement entropy and panel (c) the one of the global fidelity, panel (d) reports its time average. In panel (b), the inset reports the dynamics of $S(t)$ with both entanglement entropy and times rescaled by $L$. In panel (c), the global fidelity is compared with $\prod_j \mathcal{F}_j$ (gray line) for $L=20$.
Panels (e),(g) and (h) report dynamics, time average and standard deviation of the local fidelity $\mathcal{F}_{1-\textrm{site}}$ for different $L$, panel (f) reports the fidelity over block of variable size with $L=20$. Panel (i) reports the dynamics of the fluctuation of the $Z$-magnetization with respect to its initial value in each site for $L=20$.}
\label{fig:sketch_blockaded}
\end{figure}
The structure of the blockaded state reflects the constraints of the Hilbert space in which it is embedded, indeed, it exhibits a perfect Fibonacci-like structure, that is (see appendix~\ref{app:fibo})
\begin{equation}
    \ket{\varphi_{L}} = \sqrt{\frac{F_{L+1}}{F_{L+2}}} \ket{0, \varphi_{L-1}} + \sqrt{\frac{F_{L}}{F_{L+2}}} \ket{1,0,\varphi_{L-2}},
    \label{eq:fib}
\end{equation}
where we have introduced the compact notation $\ket{0}\otimes\ket{\varphi_{L-1}}=\ket{0, \varphi_{L-1}}$ and $\ket{1}\otimes\ket{0}\otimes\ket{\varphi_{L-2}}=\ket{1,0,\varphi_{L-2}}$. 
Using Eq.~\eqref{eq:fib} and other similar ways of rewriting the state, one can have access the all the single-site reduced density matrices that read (see appendix~\ref{app:fibo})
\begin{align}
    \rho_j^{(\varphi_L)}=&\frac{F_{j+1}F_{L-j+2}}{F_{L+2}} \ket{0}\bra{0} +\nonumber \\& \frac{F_{j}F_{L-j+1}}{F_{L+2}}(\ket{0}\bra{1} + \ket{1}\bra{0} + \ket{1}\bra{1}),
    \label{eq:density_matrix}
\end{align}
where $\rho^{(\varphi_L)}_j$ denotes the partial trace of $\ket{\varphi_L}\bra{\varphi_L}$ over the entire chain, excluding the site $j$.

We first analyze the the overlap of this state with the spectrum of the system, see Fig.~\ref{fig:sketch_blockaded}(a). In analogy with the behavior observed in the other states, the overlap increases exponentially with energy. However, unlike the other cases, here we observe the emergence of a dominant contribution from a single state, namely, the highest excited state $\ket{E_{n_{\rm max}}}$, which significantly outweighs the influence of the others and governs the overall behavior. Thus, the state is highly localized in the eigenstates basis. Although its dynamical behavior is also shaped by the significant contribution of numerous other eigenstates with non-zero overlap. This results in a complex evolution that departs from ideal scarred dynamics and gives rise to a rich dynamical structure that warrants a more in-depth exploration.

Our systematic analysis passes then through the dynamics of entanglement entropy and global fidelity, compared with $\prod_j \mathcal{F}_j(t)$, Figs.~\ref{fig:sketch_blockaded}(b),(c). The entanglement entropy dynamics remains relatively low compared to that observed during the ergodic dynamics of the homogeneous state $\ket{\boldsymbol{0}}$. Nevertheless, it exhibits a clear dependence on system size, with the entanglement increasing as the size grows. The overall behavior is distinctly non-ergodic: the entanglement entropy oscillates in time without rapidly reaching a stationary value. However, its strong dependence on $L$, which is well captured by the volume-law scaling $S(t, L) \sim L\, f\left(\frac{t}{L}\right)$ (see inset), indicates that, despite being dominated by the highest-energy scarred state, the dynamics of the blockaded state is significantly influenced by the multitude of other eigenstates with smaller overlap.

At the system sizes considered, due to the large overlap with a specific eigenstate, the global fidelity never vanishes; instead, it oscillates around values significantly different from zero. This indicates that the system retains a strong memory of its initial configuration throughout the evolution. Notably, the product of the local fidelities, illustrated here for $L=20$, closely follows and largely dominates the global fidelity. This observation suggests that local properties of the system are remarkably well preserved, pointing to the presence of stable local reminiscence.

Given the strongly oscillatory and seemingly erratic behavior of the global fidelity over time, we focus our analysis on its time-averaged value $\overline{\mathcal{F}}_t \coloneqq\frac{1}{t}\int_0^t \mathcal{F}(t') \, dt'$. This approach allows us to smooth out the fluctuations and extract more meaningful information about the system's long-time dynamics. In particular, certain properties of the global fidelity, such as its scaling with system size, are more clearly revealed through this time-averaged behavior, shown in Fig.~\ref{fig:sketch_blockaded}(d). The time-averaged global fidelity exhibits an initial decay at short times, followed by stabilization around a non-zero value that decreases as the system size increases. This decrease is naturally expected, as the dimension of the Hilbert space grows with system size, causing the overlap with any specific initial state to become increasingly diluted.

Secondly, we study the single-site local fidelity averaged over all the sites, $\mathcal{F}_{\rm 1-site}$, see Fig.~\ref{fig:sketch_blockaded}(e). The latter remains predominantly close to unity throughout its time evolution, providing strong evidence for the presence of local reminiscence. Although it experiences a transient decay at short times, the fidelity subsequently recovers and returns near to its initial value. The average value of the upper bound of the trace distance at long times is $(\overline{\mathcal{D}_{1-\rm site}^{\rm max}})_{t=100}\approx 0.095$, confirming that the evolved state remains close to its initial configuration at the local level throughout the dynamics. This behavior reflects a robust attraction toward the original local configuration, highlighting the system’s tendency to preserve local properties despite overall dynamical changes. 

Important information about local reminiscence are also encoded in the local fidelity between states of subsystems defined on blocks of the form $[1, \ell]$, with variable size, which is shown in Fig.~\ref{fig:sketch_blockaded}(f). As the block size $ \ell $ of the local subsystem increases, the corresponding local fidelity inevitably decreases.
%reflecting the increasing entanglement with the rest of the system. 
However, the differences in fidelity between various values of $ \ell $ are more pronounced for small $ \ell $, while for larger $ \ell $, the fidelity dynamics become denser and less distinguishable. This indicates that local reminiscence is most prominent for small subsystems, suggesting that it is fundamentally a property of limited spatial extent. For instance, there is a clear gap in the fidelity behavior when moving from $ \ell = 1 $ to $ \ell = 2 $, highlighting that local properties are especially well preserved at the single-site level.

The quality and stability of a locally reminiscent dynamical regime are quantified by a time-averaged value of the local fidelity close to unity, accompanied by a standard deviation close to zero. This is precisely what is observed in Figs.~\ref{fig:sketch_blockaded}(g),(h). At short times, as the system slightly departs from its local initial conditions, the time-averaged local fidelity rapidly decreases, while the standard deviation increases sharply. After reaching, respectively, a minimum and a maximum, the two quantities evolve in opposite directions: the time average continues to grow and gradually relaxes toward a value close to one, whereas the standard deviation slightly decreases and stabilizes at a notably small value. Both behaviors provide strong evidence for the presence of stable local reminiscence. Remarkably, as the system size $L$ increases, the time-averaged local fidelity tends to increase, while the standard deviation tends to decrease. This trend indicates that local reminiscence not only remains robust but is even enhanced in larger systems, suggesting that local memory becomes more stable and favored with increasing global system size. This highlights the nontrivial local stability of local reminiscent states.

Finally, we analyze the dynamics of local observables by studying the fluctuations of the local magnetizations with respect to the initial condition, $|Z_j(t) - Z_j(0)|$, which are shown in Fig.~\ref{fig:sketch_blockaded}(i) for a system of size $L = 20$. The dynamics is characterized by oscillations followed by relaxation phases of small amplitude, significantly weaker than those observed in non-reminiscent cases such as $\ket{\Theta_+(\pi/4)}$. Using Eq.~\ref{eq:density_matrix} and the definition $Z_j = \ket{1}_j\bra{1} - \ket{0}_j\bra{0}$, the initial magnetization can be computed as
\begin{equation}
    Z_j(0)=\operatorname{Tr} \Bigl[ Z_j \rho_j^{(\varphi_L)}\Bigr]=\frac{F_jF_{L-j+1} - F_{j+1}F_{L-j+2}}{F_{L+2}}
\end{equation}
which, in the large-$L$ limit, can be directly expressed in terms of the golden ratio $\varphi$. For example, for $j = 1$, we find $Z_1(0) \approx \frac{1}{\varphi^2} - \frac{1}{\varphi}$, so that during the dynamics the magnetization oscillates around values directly related to the golden ratio.

\section{Asymptotic behavior and spectral properties of the system}
\label{sec:spectr}
The dynamical properties discussed in the previous sections are obviously related to the energy spectrum of the system.
The main feature we observed is that different initial states give rise to markedly different dynamical properties both at the global and local level. We start by noticing that the change in the behavior is caused solely by the different initial states: the system Hamiltonian does not change. In this section we give a rationale for such a behavior by linking the asymptotic behavior to the spectral measure induced by the initial state. The link between the dynamical properties and the spectral measures has been investigated in the case of aperiodic quantum walks~\cite{LoGullo2017}, interacting one-dimensional quantum systems in aperiodic potentials~\cite{LoGullo2020} and also in relation to the quantum chaotic systems perturbed with aperiodic forcing~\cite{Milek1990}.
The Fidelity can be written as \(\mathcal{F}(t)=|\nu(t)|^2\), where $\nu(t)$ is the survival amplitude and it is given by:
\begin{equation}
\nu(t)=\braket{\psi(0)|\psi(t)}=\sum\limits_{n=1}^{N} |c_n|^2 e^{-i E_n t},
\end{equation}
where $c_n=\braket{E_n|\psi(0)}$, $N$ is the number of states that contribute to the dynamics.

In the limit of large system size \(N\rightarrow \infty\) we can rewrite the survival amplitude as~\cite{Last1996}:

\begin{equation}
\label{eq:vacuum_ampl}
\nu(t)=\int_{\sigma(H)} d\mu_{\psi_0}(\omega) e^{-i\omega t} %\equiv \tilde\mu_0(t)
\end{equation}

where we introduced the spectral measure \( d\mu_{\psi_0}(\omega)\), the shorthand notation \(\psi_0=\psi(0)\) and the integral is intended over the spectrum \(\sigma(H)\) of the Hamiltonian. It follows that the survival amplitude $\nu(t)$ is the Fourier transform of the spectral measure.
Using the Lebesgue decomposition theorem we can decompose the measure into an absolutely continuous \(d\mu_{ac}\), singular continuous \(d\mu_{sc}\) and pure point \(d\mu_{pp}\) parts. 
This induces a decomposition of the Hilbert space into mutually orthogonal subspaces and of the spectrum \(\sigma(H)=\sigma_{ac}(H) \cup\sigma_{sc}(H)\cup \sigma_{pp}(H)\).
We can therefore rewrite the survival amplitude as
\begin{align}
\nu(t)&=\sum\limits_{\alpha}\int_{\sigma_{\alpha}(H)} d\mu_{\psi_0,\alpha}(\omega) e^{-i\omega t}.
\end{align}
We can now resort to two results from measure theory to investigate the nature of the spectral measure induced by the initial state \(\psi(0)\): the Riemann-Lebesgue (RL) lemma and the Wiener (WN) theorem~\cite{Last1996}:
\begin{itemize}
\item RL : if \(\mu\) is absolutely continuous then the Fourier transform 
\(|\nu(t)|\rightarrow 0 \text{ for } t\rightarrow \infty\)
\\
\item WN : \(\lim\limits_{t-\infty} \overline{|\nu|^2}_t  = \sum\limits_n |\mu_{0,pp}(\omega_n)|^2\)
\end{itemize}

where \(\overline{|\nu|^2}_t = t^{-1}\int_0^t dt' \, |\nu(t')|^2 = \overline{\mathcal{F}}_t\) is the time-average (Césaro average) of the fidelity introduced in the previous section.

In the following we will use these results from measure theory in the attempt of characterizing the spectral measure induced by the states considered above. Specifically the RL lemma will tell us that a decay of the survival amplitude at long times is signaling that the measure induced by the initial state is absolutely continuous. We will then resort to Wiener's theorem to check for the presence of pure-point one in those cases in which the survival amplitude does not decay to zero.

\begin{figure*}
    \centering
    \includegraphics[width=0.75\linewidth]{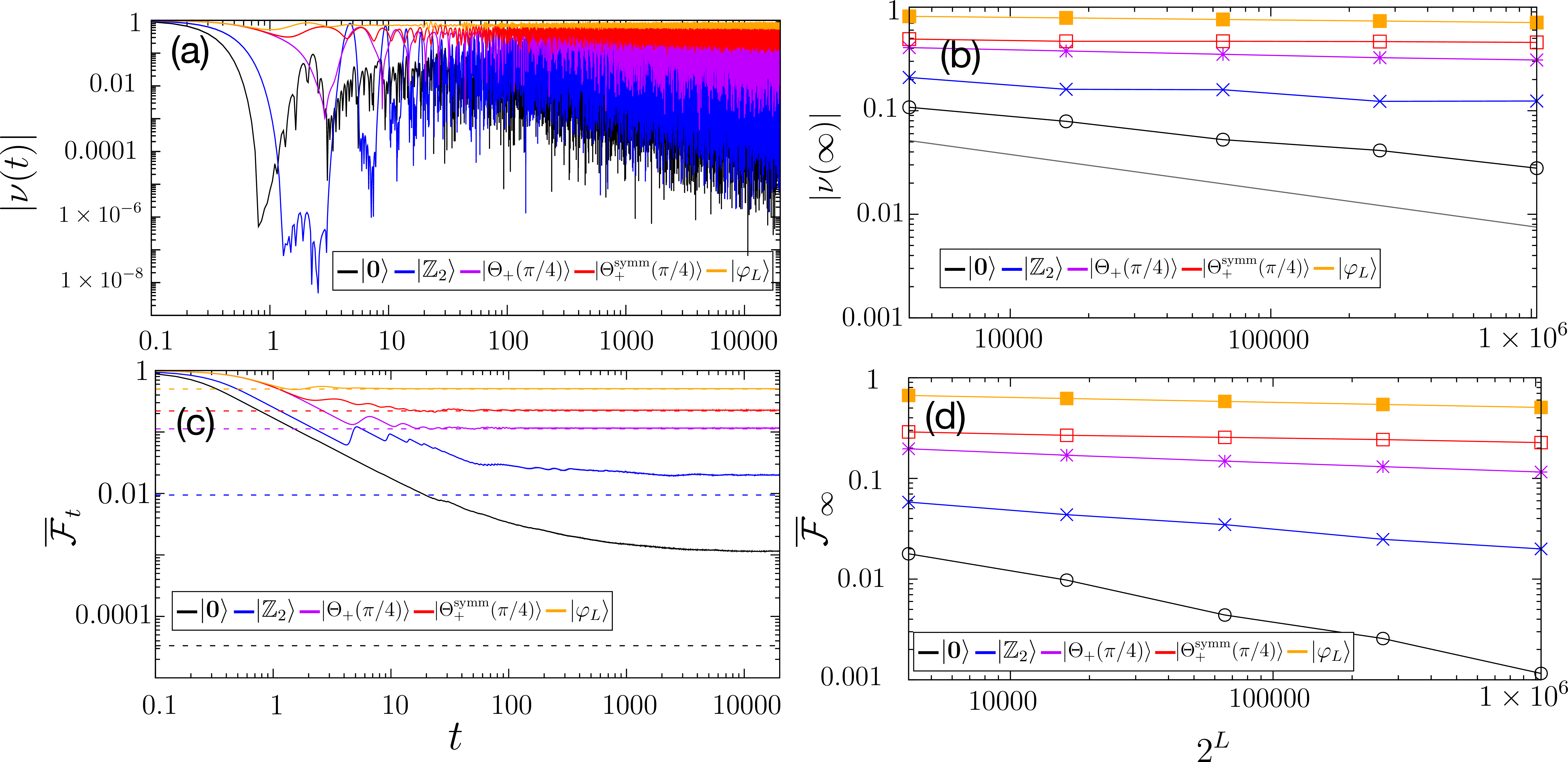}
    \caption{Fidelity at long times. In panel (a) the dynamics of $|\nu(t)|$ until long times is shown for the five different states discussed in the previous sections. For the \(\ket{\Theta}\) and \(\ket{\Theta_+^{\rm symm}}\) we chose \(\theta=\pi/4\). In panel (b) its value at long times $|\nu(\infty)|$ is extracted and reported for different values of the size of the system $L$ in log-log scale, for different initial states. To avoid fluctuations, we extract the average at long times $t=20000$ on the last $500$ times. The gray line reports $1/\sqrt{F_{L+2}}$, where $F_{L+2}$ is the dimension of the constrained Hilbert space. In panel (c) the time-average of the fidelity at long times is reported. The horizontal dashed lines represent $ \sum\limits_{n=1}^{N_s} |\braket{S_n|\psi(0)}|^4$ for any of the different initial states indicated in the legend which is a lower bound for $\overline{\mathcal{F}}_t$, or equivalently $\overline{|\nu|^2}_t$. In both panels the size of the system is $L=20$. In panel (d) the time average of the fidelity at long times $\overline{\mathcal{F}}_{\infty}$, or alternatively $\overline{|\nu|^2}_{\infty}$, for different values of $L$ and different initial states is reported, in log-log scale.}
    \label{fig:long_fidelity}
\end{figure*}

In Fig.~\ref{fig:long_fidelity} (a) we show $|\nu(t)|$ of the initial state at long times for all states discussed in the previous sections for a system with \(L=20\). For the homogeneous state, at long times, it drops to small values,  by the RL lemma we can infer that the measure induced by the homogeneous state is absolutely continuous. Stated differently, the homogeneous state \(\ket{\boldsymbol{0}}\) has a non-vanishing overlap \textit{mostly} with the absolutely continuous part \(\sigma_{ac}(H_{\rm PXP})\) of the spectrum of the Hamiltonian. Nevertheless some oscillations remain at long times, signaling the presence of other spectral components.
Things are markedly different for all other states. Starting from the $\ket{\mathbb{Z}_2}$ state we observe persistent oscillations which do not die out at long times. 

We also looked at the scaling with the system size of the long value of the survival amplitude \(|\nu(t)|\) at long times. The results are reported in Fig.\ref{fig:long_fidelity} (b). We can observe a linear scaling with the size of the Hilbert space for the homogeneous state, whereas for the other states the value is less sensitive  to the system size. This behavior reflects the nature of the eigenstates of a continuous and discrete energy spectrum respectively. In the former case one expects eigenstates which are extensive with the system size, whereas in the second case the eigenstates are less sensitive to a change in the system size.

To substantiate these statements we look at the time averaged fidelity and exploit the Wiener's theorem to find a relation to the spectral properties of the Hamiltonian. The time averaged fidelity at long times is a figure of merit that can be used to quantify the amplitude of the oscillations as well as their frequency over time.
The higher its values the more persistent these oscillations are. We used a similar approach in the previous sections to characterize the local fidelity. In Fig.~\ref{fig:long_fidelity}(c) we show the dynamics of the time-averaged fidelity, indicated as $\overline{\mathcal{F}}_t$, which corresponds to $\overline{|\nu|^2}_t$, for $L=20$, and in Fig.~\ref{fig:long_fidelity}(d) its asymptotic value $\overline{\mathcal{F}}_\infty$ as a function of $L$.
We see that for the homogeneous state, as the system size increases the average fidelity drops signaling, according to the Wiener's theorem, that the pure point part of the spectrum tends to zero in the thermodynamic limit. For the $\ket{\mathbb{Z}_2}$ state things are different, as the system size is increased, the long time average of the fidelity seems to become independent of the system size in the thermodynamic limit.
The same is true for all other states considered with the noticeable difference that going from the \(\ket{\Theta_+}\), to \(\ket{\Theta_+^{\text{symm}}}\) ad eventually to the blockaded \(\ket{\varphi_L}\) the asymptotic value sets to higher and higher values. Once again, resorting to the Wiener's theorem we can conclude that this is due to the state inducing a measure with a larger component on the pure point part of the spectrum.

We try now to identify the states which are responsible for such a behavior. 
In Ref.~\cite{choi2019emergent} it was argued that the states with maximum overlap with the \(\ket{\mathbb{Z}_2}\)
within approximately equally spaced energy bands, the QMBS, are discrete states and are responsible for the revivals in the fidelity and the deviation from ETH. 
Using these two criteria, namely minimum entropy and maximum overlap with \(\ket{\mathbb{Z}_2}\),  we have identified the QMBS states. According to Wiener's theorem we have 

\begin{equation}
%\lim\limits_{T-\infty} \langle\tilde \mu\rangle_T 
\lim\limits_{t-\infty} \overline{|\nu|^2}_t \ge \sum\limits_{n=1}^{N_s} |\braket{S_n|\psi(0)}|^4,
\label{eq:scarred}
\end{equation}
where \(\ket{S_n}\) are the scarred eigenstates of the Hamiltonian and \(N_s\) is their number.
The inequality comes from the fact that we have no guarantee that these states are the only ones populating the discrete part of the spectrum \(\sigma_{pp}(H_{\rm PXP})\).

In Fig.~\ref{fig:long_fidelity}(c) we show this limit for the different states.
In the case of the homogeneous state, which has the lowest overlap with the scars, we see that the fidelity approaches a value which is about ten times larger than that obtained by considering only the scarred eigenstates.
For all other states, where the scarred states have a larger overlap with the initial states upper bound in Eq.~\ref{eq:scarred} is approached. These findings suggest that the scarred states identified in Ref.~\cite{turner2018quantum} and widely studied since, might belong to the pure point spectrum of \(H_\text{\rm PXP}\). To confirm these suppositions, simulations with larger system sizes are necessary, especially in light of the findings reported in Ref.~\cite{turner2018quantum} in which the authors have shown that at sizes \(L\ge 34\) the scars might hybridize with states having a volume scaling.

Instead we can argue that the local reminiscence we introduced emerges from the scarred states. In fact, both of the two states exhibiting this behavior show large values for $\lim\limits_{t-\infty} \overline{|\nu|^2}_t$, which suggests that the spectral regions containing the scar states play a significant role in their dynamics.

\section{Conclusions $\&$ outlook}
\label{sec:concl}
We have explored the dynamics of the PXP model through the concept of local reminiscence, defined as the tendency of reduced density matrices on small subsystems to remain close to their initial configuration over time, even as the system undergoes significant global evolution. This reflects the stable retention of local information despite the spread of quantum information at the global level. While global indicators such as fidelity and the entanglement entropy between the two halves of the chain are commonly used to characterize ergodicity or quantum scarring, we have shown that local fidelities offer a more sensitive probe of information retention at the subsystem level. Unlike global measures that average over the entire system, local fidelities capture the evolution of specific regions, revealing how local memory persists or decays. Our results demonstrate that local fidelities serve as a powerful diagnostic of local memory effects that may be hidden in global quantities. By tracking the temporal evolution of reduced density matrices, we uncover subtle features of the dynamics and highlight the intricate interplay between local and global behavior in the PXP model, offering deeper insight into memory retention in quantum many-body systems.

In this work, we have analyzed three classes of initial states in the PXP model: $\ket{\Theta_+}$, $\ket{\Theta_+^{\rm symm}}$, and $\ket{\varphi_L}$. The state $\ket{\Theta_+}$ does not exhibit local reminiscence, as local subsystems quickly lose memory of the initial configuration. In contrast, the symmetric state $\ket{\Theta_+^{\rm symm}}$, that is a superposition of $\ket{\Theta_+}$ and its spatial reflection, shows strong local reminiscence, with local fidelities remaining close to one throughout the evolution. The time-averaged fidelity and its fluctuations respectively increase and decrease with system size, suggesting growing stability in the thermodynamic limit. Local reminiscence occurs regardless of boundary conditions, but PBC enhance it by better preserving local information and reducing edge-induced distortions. Finally, the blockaded state $\ket{\varphi_L}$, a superposition of all the computational basis states without two or more adjacent excitations and structured by the Fibonacci sequence, also displays robust local reminiscence. Its local fidelities remain high over time, with decreasing fluctuations as system size increases, further confirming the persistence of local memory in highly constrained Hilbert spaces.

We related the emergence of local reminiscence to the spectral properties of the Hamiltonian of the system. Specifically, we have found evidences suggesting that the previously identified scarred eigenstates, namely those with lowest entropy and highest overlap with the \(\ket{\mathbb{Z}_2}\) might belong to the discrete part of the spectrum, although simulations with larger system sizes are required to confirm this hypothesis. Instead, we can say that having a large overlap with the scarred eigenstates results in a good local reminiscence of the system.

The emergence of local reminiscence from specific initial states in the PXP model opens new avenues for exploring this phenomenon in other quantum many-body systems. A key direction for future work is identifying the spectral and microscopic features responsible for the onset and stability of local reminiscence, with the goal of uncovering universal mechanisms behind it. Additionally, deviations from the idealized initial states considered in this study can be systematically investigated to assess the robustness of local reminiscence under perturbations. It is also compelling to explore potential connections between local reminiscence and other phenomena in quantum many-body systems, such as many-body localization~\cite{alet2018many, pal2010many, nandkishore2015many, abanin2019colloquium, sierant2025many}  and metastability~\cite{yin2025theory}. Understanding these relationships may provide deeper insights into the rich landscape of non-thermalizing dynamics in closed quantum systems.

\section{Acknowledgments}
The authors thank Prof. Li You for valuable insights on the Rydberg atoms setups.
NLG thanks all the participants to the first edition os the ``Quantum coast to coast" conference (N. Defenu, G. Gori, C. Paletta, A. Trombettoni, and P. Vignolo) for fruitful discussions.
This work was partially funded by the
PNRR MUR Project No. PE0000023-NQSTI through the
secondary projects QuCADD, ThAnQ and QuSOE. We acknowledge the use of the Finnish CSC facilities under the project 2010295 ``Quantum Reservoir Computing".

\appendix

\section{Idea behind the derivation of the PXP Model for a 1D Rydberg atom chain}
\label{app:details_derivation}
The starting point for this derivation is the complete many-body Hamiltonian describing an array of $L$ Rydberg atoms under coherent laser driving. The full Hamiltonian for a one-dimensional chain of driven Rydberg atoms incorporates three distinct physical contributions: the coherent laser driving that couples ground and Rydberg states (defined as $\ket{0}$ and $\ket{1}$), the detuning that controls the relative energy of Rydberg excitations, and the long-range van der Waals interactions between Rydberg atoms. The Hamiltonian reads~\cite{saffman2010quantum,browaeys2020many,henriet2020quantum,wu2021concise}
\begin{align}
    H&= \sum_j H^{(j)}_{\text{drive}}+\sum_{i<j}H_{\text{int}}^{(ij)}\\
    &=\sum_{j=i}^L \left[ \frac{ \Omega}{2}X_j+\Delta n_j \right]+\sum_{i<j}\frac{C_6}{|r_i-r_j|^6}n_i n_,\nonumber
\end{align}
where $\Omega$ is the Rabi frequency, $\Delta$ is the detuning, $r_j=a\,j$ indicates the position of the $j$-th atom in the array, $a$ is the lattice spacing and the $C_6$-coefficient describes the intensity of the van der Waals interaction between two Rydberg atoms. We have also introduced the operator $n_j=\frac{1}{2}(1+Z_j)$, which is the projector onto the $\ket{1}_j$ state. 

This Hamiltonian embodies the complete description of the driven Rydberg system, including all relevant energy scales and interaction mechanisms. However, the presence of long-range interactions with varying distance dependencies makes this Hamiltonian challenging to analyse directly, both analytically and numerically. The key insight that enables significant simplification is the recognition that in the strong blockade regime~\cite{pizzi2025genuine}, certain configurations become energetically inaccessible and can be projected out of the dynamics, leading to a dynamics that is effectively described by the PXP model.

The derivation of the PXP model relies on a clear separation of energy scales in the system. The strong blockade regime~\cite{lesanovsky2012interacting,browaeys2020many} is characterized by the condition that the nearest-neighbor van der Waals interaction energy substantially exceeds both the Rabi frequency and the detuning: $\frac{C_6}{a^6} \gg  \Omega, \, |\Delta|$. Under this condition, any attempt to create two adjacent Rydberg excitations results in a large energy penalty that places such configurations far outside the spectral bandwidth of the driving laser. The hierarchical energy scale separation enables a systematic projection of the dynamics onto the blockade-constrained subspace. States that violate the blockade constraint acquire large energy shifts and become effectively decoupled from the low-energy dynamics. 

In this limit, the system is effectively described by the PXP Hamiltonian, that reads~\cite{turner2018quantum,ho2019periodic}:
\begin{equation}
    H_{\rm PXP} = \frac{\Omega}{2}\sum_{j=1}^{L} P_{j-1} X_{i} P_{j+1} + \Delta \sum_{j=1}^{L} n_j,
\end{equation}
where $P_j \equiv 1-n_j =  \ket{0}_j\bra{0}$ is the projector onto the de-excited $\ket{0}$ state.

In the case of open boundary conditions (OBC), the Hamiltonian is 
\begin{equation}
    H_{\rm PXP} = \frac{\Omega}{2} \Bigl(X_1 P_2 + \sum_{j=2}^L P_{j-1}X_j P_{j+1} + P_{L-1}X_{L}\Bigr) + \Delta \sum_{j=1}^{L} n_j.
\end{equation}

This remarkably simple effective Hamiltonian captures the essential physics of driven Rydberg arrays in the strong blockade regime. The first term represents the kinetic energy associated with coherent excitation, subject to the constraint that such hopping can only occur when neighboring sites are unoccupied. In fact, the allowed flip operator is
\begin{equation}
    P_{j-1}X_j P_{j+1},
\end{equation}
and acts as follows:
\begin{itemize}
    \item [1.] If either neighbor is $\ket{1}$, the projector kills the amplitude and the flip does not occur.
    \item [2.] Only when both neighbors are $\ket{0}$ does $X_j$ act normally. 
\end{itemize}
The second term provides a chemical potential for Rydberg excitations, effectively controlling the energy cost of creating excitations.
\section{Scaling of the product of local fidelities}
\label{app:scaling_prod}
\begin{figure}[!t]
\centering
\includegraphics[width=\linewidth]{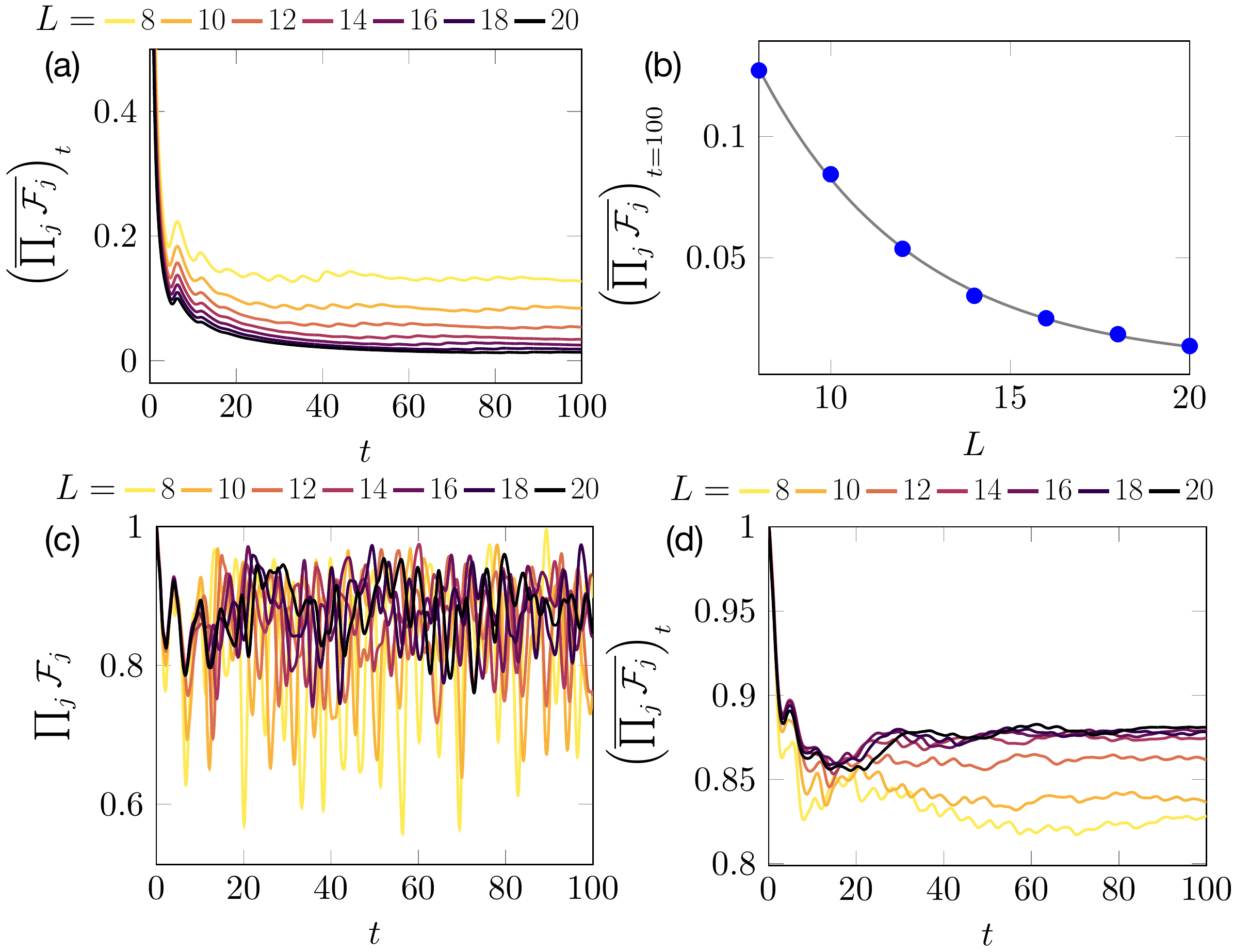}
\caption{Panel (a): time average of the product of the local fidelities $\left(\overline{\prod_j\mathcal{F}_j}\right)_t$ for different values of the size with the system initialized in $\ket{\Theta_+(\pi/4)}$. Panel (b): scaling in size of $\left(\overline{\prod_j\mathcal{F}_j}\right)_t$ at $t=100$ (blue points) with the system initialized in $\ket{\Theta_+(\pi/4)}$. data are well fitted by an exponential curve $a \exp(-b(L-l)) + c$ with $a=0.63389, \, b = 0.232268,\, l=0.909499,\, c=0.005807$ (black line). Panels (c)-(d): dynamics of the product of local fidelities for different values of the size $L$, when the system is initialized in $\ket{\Theta^{\rm symm}_+(\pi/4)}$. Panel (c) reports $\prod_j \mathcal{F}_j(t)$, while panel (d) reports its time average $\left(\overline{\prod_j\mathcal{F}_j}\right)_t$.}
\label{figapp:product}
\end{figure}
In this section, we study the scaling in size of the product of local fidelities $\prod_{j=1}^L\mathcal{F}_j(t)$ and its time-average
\begin{equation}
\left(\overline{\prod_j\mathcal{F}_j}\right)_t=\frac{1}{t}\int_0^t dt' \prod_{j=1}^L\mathcal{F}_j(t');  
\end{equation}
We consider the states $\ket{\Theta_+(\pi/4)}$ and $\ket{\Theta^{\rm symm}_+(\pi/4)}$. For the former, we analyze the behavior of $\left(\overline{\prod_j\mathcal{F}_j}\right)_t$ for different system sizes, as shown in Fig.~\ref{figapp:product}(a),(b), since the quantity $\prod_{j=1}^L\mathcal{F}_j(t)$ is already reported in Fig.~\ref{fig:local_theta+}(c). We observe that the quantity under investigation decays both in time and with increasing size; in particular, its scaling with $L$ is well captured by an exponential decay, indicating that local memory is progressively lost in the large–system-size limit.

In Fig.~\ref{figapp:product}(c),(d), we consider the local reminiscent state $\ket{\Theta^{\rm symm}_+(\pi/4)}$. We analyze both $\prod_{j=1}^L\mathcal{F}_j(t)$ (panel (c)) and its time average (panel (d)). The former exhibits a rather erratic and non-monotonic behavior in time, but despite these fluctuations the overall trend is that the product approaches unity as the system size increases. This tendency becomes even more evident when examining the time-averaged quantity, which smooths out oscillations and highlights the underlying scaling. As $L$ increases, the time-averaged product of local fidelities converges from below toward a value close to $0.9$, indicating a substantial degree of local stability and reminiscence. Crucially, the scaling with $L$ is far from exponentially suppressed, in stark contrast with what one expects for uncorrelated product states and with non-reminiscent initial configurations such as $\ket{\Theta_+(\pi/4)}$. Instead, the observed saturation shows that correlations in $\ket{\Theta^{\rm symm}_+(\pi/4)}$ efficiently preserve its local reminiscent nature even at large sizes. This confirms that the correlated structure of the state plays a key role in maintaining robust local properties as $L$ grows.
\section{Local reminiscence in the presence of PBC}
\label{app:pbc}
\begin{figure}[!t]
\centering
\includegraphics[width=\linewidth]{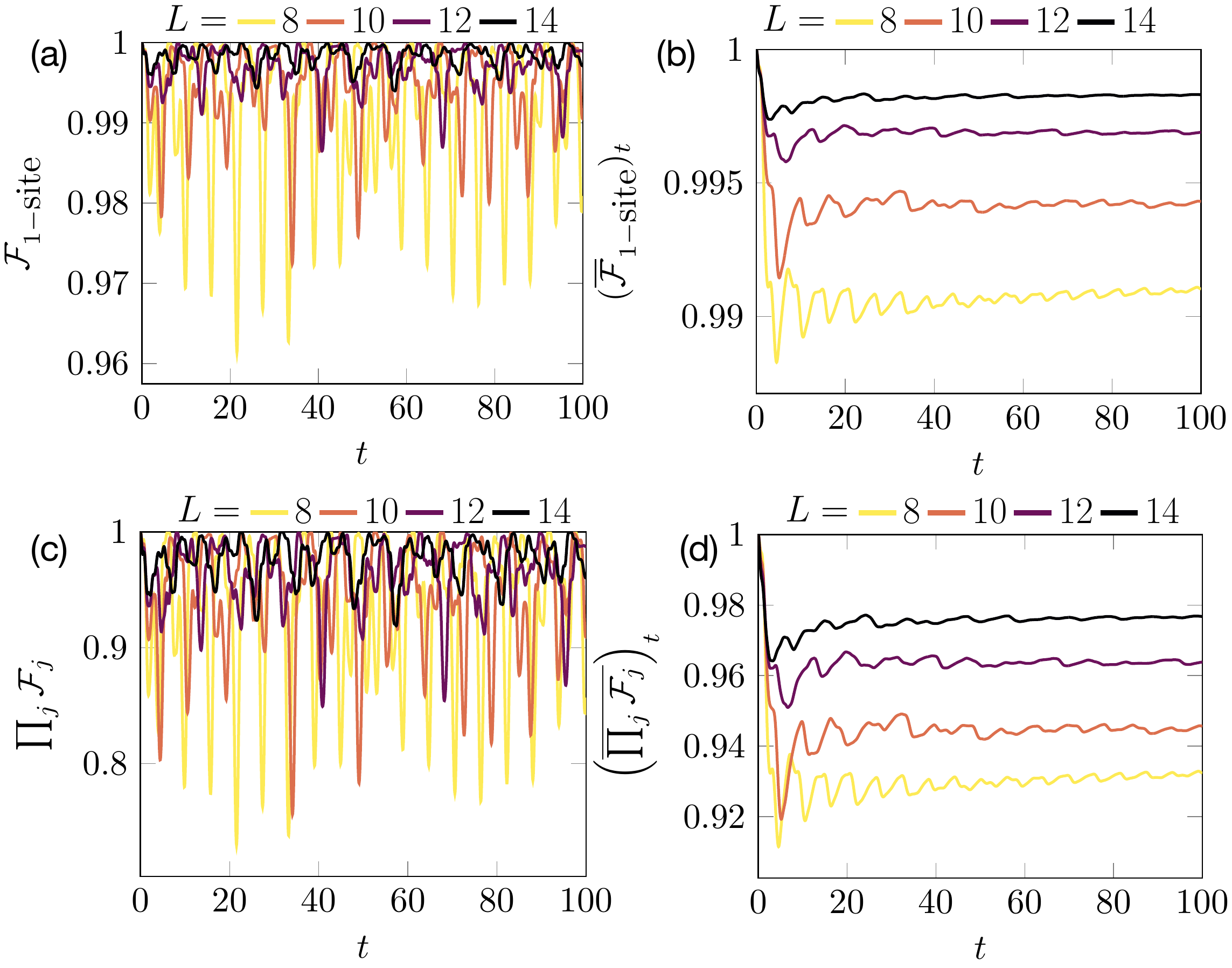}
\caption{Local fidelity analysis with PBC for the $\ket{\Theta_+^{\rm symm}(\pi/4)}$ state. Panels (a) and (b) report the dynamics of the one-site local fidelity (panel (a)) and the one of its time-average (panel (b)). Panels (c) and (d) report the product of the local fidelities, its dynamics in panel (c) and the dynamics of its time-average in panel (d). Data are reported for different values of $L$.}
\label{figapp:pbc}
\end{figure}
In this section, we analyze the properties of the local reminiscent state $
\ket{\Theta^{\rm symm}_+(\pi/4)}$ in the presence of periodic boundary conditions (PBC). 
The Hamiltonian of the system now takes the form
\begin{equation}
    H_{\rm PXP}
    = P_L X_1 P_2
    + \sum_{j=2}^{L-1} P_{j-1} X_j P_{j+1}
    + P_{L-1} X_{L} P_1,
\end{equation}
where we have fixed the driving amplitude to $\Omega/2 = 1$. Our goal is to address the following questions: (i) does local reminiscence persist in the presence of PBC? (ii) If so, is it enhanced or suppressed compared to the OBC case? To answer these questions, we study the evolution of the one-site fidelity $\mathcal{F}_{1\text{-site}}(t)$. 
Unlike in the OBC setting, here both the Hamiltonian and the initial state 
$\ket{\Theta^{\rm symm}_+(\pi/4)}$ 
are fully translationally invariant. 
As a consequence, all local reduced density matrices are identical, and the fidelity $\mathcal{F}_{1\text{-site}}(t)$ is strictly site-independent. 
Therefore, the time-averaged product of fidelities simplifies to $\left(\overline{\prod_j \mathcal{F}_j}\right)_t
= \big(\mathcal{F}_{1\text{-site}}(t)\big)^L$.

Figure~\ref{figapp:pbc} shows the time evolution of both the one-site fidelity 
$\mathcal{F}_{1\text{-site}}(t)$ and their product $\prod_j \mathcal{F}_j(t)$,
together with their corresponding time averages. 
Several features clearly emerge from the PBC data. First, the one-site fidelity remains significant throughout the dynamics and displays values that are fully comparable with those observed in the OBC case associated with local reminiscent behavior. 
Moreover, its time-averaged value increases systematically with the system size $L$, indicating that local reminiscence becomes progressively more robust as the chain grows. 
This scaling is particularly transparent under PBC, where translational invariance forces $\mathcal{F}_{1\text{-site}}(t)$ to be identical on every site, thus reinforcing the uniformity of the local memory retained by the system. Second, the magnitude of both $\mathcal{F}_{1\text{-site}}(t)$ and its chain product is visibly larger than in the corresponding OBC results. 
This enhancement suggests that closing the boundaries stabilizes the local structure of the initial state, effectively reducing edge-induced distortions that are otherwise present under open boundary conditions. 
In other words, the periodic geometry strengthens the ability of the system to preserve local information about the initial configuration, leading to a more pronounced and longer-lived form of local reminiscence. Overall, these observations indicate that periodic boundary conditions not only preserve the qualitative features of local reminiscent dynamics, but actually amplify them, making the state $\ket{\Theta^{\rm symm}_+(\pi/4)}$ even more locally stable under PXP evolution.

\section{Fibonacci structure of the blockaded state}
\label{app:fibo}
The blockaded state on a chain composed of $L$ sites is defined as
\begin{equation}
\ket{\varphi_L} = \frac{1}{\sqrt{F_{L+2}}} 
\sum_{\ket{\boldsymbol{s}} \in \mathcal{B}_L} \ket{\boldsymbol{s}},
\label{eq:app_blockaded}
\end{equation}
where $F_{L+2}=\dim(\mathcal{B}_L)$ is the dimension of the constrained Hilbert space $\mathcal{B}_L$, $\ket{\boldsymbol{s}}=\ket{s_1,s_2,\ldots,s_L}$ are the states belonging to the computational basis $\mathcal{B}_L$ that satisfy the blockade constraint on nearest-neighbor excitations. From now on, we introduce the notation $\ket{\varphi_{l}}_{[m,m+l-1]}$ that indicates a blockaded state defined on the block of sites ranging from the site $m$ to the site $m+l-1$ which is composed of $l$ sites. Following this notation, the state in Eq.~\eqref{eq:app_blockaded} can be written as $\ket{\varphi_L}_{[1,L]}$. We now observe that any basis state $\ket{\boldsymbol{s}}$ can have the first site in either $\ket{0}$ or $\ket{1}$. Considering open boundary conditions (OBC), the sum over all possible computational basis states $\ket{\boldsymbol{s}} \in \mathcal{B}_L$ that satisfy the Rydberg blockade constraint takes the recursive form:
\begin{align}
    \ket{\xi_L}_{[1,L]}\coloneqq\sum_{\ket{\boldsymbol{s}} \in \mathcal{B}_L} \ket{\boldsymbol{s}} = & \ket{0}_1 \otimes \ket{\xi_{L-1}}_{[2,L]} +\\& \ket{1}_1 \otimes \ket{0}_{2}\otimes\ket{\xi_{L-2}}_{[3,L]}.
\end{align}
This decomposition follows from the fact that if the first site is in the de-excited state $\ket{0}$, there is no constraint on the second site, and the remaining $L-1$ sites can form any superposition of basis states obeying the blockade constraint. Conversely, if the first site is in the excited state $\ket{1}$, the blockade enforces the second site to be in $\ket{0}$, and the constraint then applies recursively to the remaining $L-2$ sites. Since $\ket{\xi_L}=\sqrt{F_{L+2}}\ket{\varphi_L}$, we obtain
\begin{equation}
    \ket{\varphi_{L}} =\underbrace{\underbrace{\sqrt{\frac{F_{L+1}}{F_{L+2}}}\ket{0, \varphi_{L-1}}}_{\text{sum of $F_{L+1}$ states}}+\underbrace{\sqrt{\frac{F_{L+1}}{F_{L+2}}}\ket{1,0,\varphi_{L-2}}}_{\text{sum of $F_L$ states}} }_{\text{sum of $F_{L+2}$ states}},
    \label{eq:fib_structure_exact}
\end{equation}
where, for clarity, 
\begin{align}
&\ket{0, \varphi_{L-1}}\coloneqq\ket{0}_1 \otimes \ket{\varphi_{L-1}}_{[2,L]} , \\& \ket{1,0,\varphi_{L-2}}\coloneqq\ket{1}_1\otimes \ket{0}_2 \otimes \ket{\varphi_{L-2}}_{[3,L]}.
\end{align}
We now recall that, for large $L$, the Fibonacci number satisfies $F_L \approx \frac{\varphi^{L+1}}{\sqrt{5}}$, where $\varphi = \frac{1+\sqrt{5}}{2}$ is the golden ratio. Using this approximation, Eq.~\eqref{eq:fib_structure_exact} reduces to
\begin{equation}
    \ket{\varphi_{L}} \approx \frac{1}{\sqrt{\varphi}}\ket{0, \varphi_{L-1}}+\frac{1}{\varphi}\ket{1,0,\varphi_{L-2}},
\end{equation}
highlighting the elegant Fibonacci structure of $\ket{\varphi_L}$.

We now turn to the local features of the system. Let us compute the single-site density matrix on $j=1$:
\begin{align}
\begin{split}
    \rho_1^{(\varphi_L)}=&\operatorname{Tr}_{[2,L]}\left[\ket{\varphi_L}_{[1,L]}\bra{\varphi_L}\right]=\\
    &\frac{F_{L+1}}{F_{L+2}}\ket{0}_1\bra{0}\operatorname{Tr}_{[2,L]}\left[\ket{\varphi_{L-1}}_{[2,L]}\bra{\varphi_{L-1}}\right] +\\& \sqrt{\frac{F_{L+1}}{F_{L+2}}\frac{F_{L}}{F_{L+2}}}\Bigl( \ket{0}_1\bra{1}\operatorname{Tr}_{[2,L]}\left[\ket{\varphi_{L-1}}_{[2,L]}\bra{0,\varphi_{L-2}}\right]+\\
    &\ket{1}_1\bra{0}\operatorname{Tr}_{[2,L]}\left[\ket{0,\varphi_{L-2}}_{[2,L]}\bra{\varphi_{L-1}}\right]\Bigr) +\\& \frac{F_{L}}{F_{L+2}} \ket{1}_1\bra{1}\operatorname{Tr}_{[2,L]}\left[\ket{0,\varphi_{L-2}}_{[2,L]}\bra{0,\varphi_{L-2}}\right].
\end{split}
\end{align}
The traces of the first and last term are one, while 
\begin{align}
\begin{split}
    &\operatorname{Tr}_{[2,L]} \left[\ket{\varphi_{L-1}}_{[2,L]}\bra{0,\varphi_{L-2}} \right]=\sqrt{\frac{F_L}{F_{L+1}}},\\
    &\operatorname{Tr}_{[2,\dots,L]} \left[\ket{0,\varphi_{L-2}}_{[2,L]}\bra{\varphi_{L-1}}\right]=\sqrt{\frac{F_L}{F_{L+1}}};
\end{split}
\end{align}
the resulting single-site density matrix is
\begin{equation}
\rho_1^{(\varphi_L)}=\frac{F_{L+1}}{F_{L+2}}\ket{0}\bra{0}+\frac{F_L}{F_{L+2}}(\ket{0}\bra{1}+\ket{1}\bra{0})+\frac{F_L}{F_{L+2}}\ket{1}\bra{1},
\label{rho_1}
\end{equation}
that, for large $L$, reduces to
\begin{equation}
\rho_1^{(\varphi_L)} \approx \frac{1}{\varphi} \ket{0}\bra{0} + \frac{1}{\varphi^2} \left(\ket{0}\bra{1} + \ket{1}\bra{0} + \ket{1}\bra{1}\right).
\end{equation}
In the notation $\rho_j^{(\varphi_l)}$, $l$ indicates the total number of sites and $j$ the site on which we compute the reduced density matrix, the other sites are traced out.

Using this type of approach, one can find the generic form of the single-site reduced density matrix, that reads
\begin{align}
\rho_j^{(\varphi_L)}=&\frac{F_{j+1}F_{L-j+2}}{F_{L+2}} \ket{0}\bra{0} + \nonumber\\& \frac{F_{j}F_{L-j+1}}{F_{L+2}}(\ket{0}\bra{1} + \ket{1}\bra{0} + \ket{1}\bra{1}).
\end{align}

\end{document}